\documentclass{article}

\usepackage{arxiv}
\usepackage[utf8]{inputenc} 
\usepackage[T1]{fontenc}    
\usepackage{hyperref}       
\usepackage{booktabs}       
\usepackage{amsfonts}       
\usepackage{nicefrac}       
\usepackage{microtype}      
\usepackage{graphicx}
\usepackage{amssymb}
\usepackage{amsmath}
\usepackage{multirow}
\usepackage[square,numbers]{natbib}
\usepackage[perpage]{footmisc}
\usepackage{array}
\usepackage[ruled,vlined]{algorithm2e}

\title{CoV-ABM: A stochastic discrete-event agent-based framework to simulate spatiotemporal dynamics of COVID-19}

\author{
  Masoud Jalayer $^{1}$ \thanks{Corresponding author} \\
  \texttt{masoud.jalayer@polimi.it} \\
   \And
 Carlotta Orsenigo $^{1}$ \\
  \texttt{carlotta.orsenigo@polimi.it} \\
  \And
  Carlo Vercellis $^{1}$ \\
  \texttt{carlo.vercellis@polimi.it} \\
  \\
$^{1}$ Department of Management, Economics and Industrial Engineering\\ Politecnico di Milano, via Lambruschini 4B, 20156, Milan, MI, Italy
}

\begin{document}
\maketitle
\begin{abstract}
The paper develops a stochastic Agent-Based Model (ABM) mimicking the spread of infectious diseases in geographical domains. The model is designed to simulate the spatiotemporal spread of SARS-CoV2 disease, known as COVID-19. Our SARS-CoV2-based ABM framework (CoV-ABM) simulates the spread at any geographical scale, ranging from a village to a country and considers unique characteristics of SARS-CoV2 viruses such as its persistence in the environment. Therefore, unlike other simulators, CoV-ABM computes the density of active viruses inside each location space to get the virus transmission probability for each agent. It also uses the local census and health data to create health and risk factor profiles for each individual. The proposed model relies on a flexible timestamp scale to optimize the computational speed and the level of detail. In our framework each agent represents a person interacting with the surrounding space and other adjacent agents inside the same space. Moreover, families’ stochastic daily tasks are formulated to get tracked by the corresponding family members. The model also formulates the possibility of meetings for each subset of friendships and relatives. The main aim of the proposed framework is threefold: to illustrate the dynamics of SARS-CoV diseases, to identify places which have a higher probability to become infection hubs and to provide a decision-support system to design efficient interventions in order to fight against pandemics. The framework employs SEIHRD dynamics of viral diseases with different intervention scenarios. The paper simulates the spread of COVID-19 in the State of Delaware, United States, with near one million stochastic agents. The results achieved over a period of 15 weeks with a timestamp of 1 hour show which places become the hubs of infection. The paper also illustrates how hospitals get overwhelmed as the outbreak reaches its pick.
\end{abstract}


\section{Introduction}
\label{sec:s1}
The COVID-19 pandemic is growing every day across the planet and, apart from its impact on the global economy, it is causing a dramatic loss of lives. Since its first report on December 31st, 2019 in Wuhan, China, it has speedily reached 213 countries. As of today, over 7 million people have been diagnosed positive with over 400.000 casualties \cite{Worldometers2020a}. Policymakers and scholars all over the globe are under pressure to make effective decisions with the purpose of controlling and mitigating the outbreak in the short run, and of designing valid actions to contrast future possible upsurges.\\
Some non-pharmaceutical interventions have being carried out by governments and people aiming at “flattening the curve”, e.g. home quarantine, intensive hand hygiene, closure of schools and shops, prohibition of mass events \cite{Kantner2020}. However, these counter-measures caused controversies in many countries; as an example, several protests arose in the US against the closure of businesses and stay-at-home impositions \cite{Dyer2020}. One of the aims of the present work is to illustrate how these provisions can effectively flatten the curve of infected people and save lives.\\
Since 1920s many stochastic epidemic models have been proposed using differential equations of SIR (Susceptible-Infectious-Recovered) and SEIR (Susceptible-Exposed-Infectious-Recovered) to study the dynamics of diseases on populations \cite{Keeling2011}. Due to their generalized nature, however, differential equations cannot explain the effect on the outbreaks of different scenarios, geographical features or population age distribution \cite{Hunter2017}. Hence, during the past two decades computer simulations and agent-based models were proposed to address the structure and dynamics of transmission networks \cite{Squazzoni2020}. Some studies considered static graphs as the main platforms of the disease diffusion \cite{Wang2017}, and employed graph features and algorithms to identify the influential nodes \cite{Jalayer2018a}. Other studies resorted, instead, to spatial models for investigating such phenomena. As an example, Balcan et al. \cite{Balcan2010} developed a discrete stochastic epidemic model based on gridded population at the worldwide scale, whereas Cliff et al. \cite{Cliff2018} built a spatiotemporal simulator to analyse the spread of influenza in Australia using census data.\\
Some works addressing the new outbreak, SARS-CoV2, have been recently proposed. Kai et al. \cite{Kai2020} developed two SEIR models to predict the effect on the spread of the virus of wearing face masks in the society. The first model uses stochastic dynamic networks to simulate the transmission among the individuals, represented by the nodes of the network. The second model, instead, employs an agent-based framework which indicates that “the early universal masking” policy can effectively mitigate the pandemic. Cuevas \cite{Cuevas2020} proposed an agent-based simulator to mimic the spatiotemporal transmission process of COVID-19. In this work, each agent has a health profile and social characteristics, and makes decisions about its movements and interactions with other agents inside a square, which represents a generic geographical area. The results of this study show that simulations provide useful information to base the decisions aimed at mitigating the transmission risks of COVID-19 within the facilities. In \cite{Ivorra2020} Ivorra et al. developed a mathematical model to study the importance of undetected infections. They proposed a modified SEIR model, called $\theta$-SEIHRD, where $\theta$ implies the fraction of detected cases over the actual total number of infectious cases, and H denotes the state of the people who are hospitalized or self-quarantined. The authors compared their differential equation results with the adjusted number of COVID-19 cases in China, and analysed some different scenarios consequently. In another study \cite{DOrazio2020}, the authors developed an agent-based model to simulate the spread of COVID-19 based on the movements of the agents in closed built environments, such as buildings, and studied some different social-distancing scenarios. The model there proposed is calibrated on experimental data collected from the Diamond Princess cruise, and relies on two factors to determining if the virus transmits: ‌the distance between the virus carrier and the susceptible agents, and the duration of the exposure.\\
In this paper, we propose an agent-based model which simulates agents capable of working, resting, taking part in social events, meeting friends and visiting relatives stochastically. The model uses census data referred to the working context and to the health risk factors in a given geographical area, to create a realistic society. Moreover, it simulates the placement of facilities and buildings in that area by using Geographic Information System (GIS) tools to retrieve real-world geographical data, and associates them to agents who live, work or simply visit that locations. In our work distinctive traits of SARS-CoV2 are also considered; in particular, our model assumes that the virus persists inside the buildings or in outdoor areas for rather long time periods, and stochastically infects the agents in these places.\\
The main aim of the proposed framework is threefold. First, to enable a detailed description of the dynamics of the SARS-CoV2 spread. Second, to identify the high-risk places that accelerate the outbreak. Third, to provide a decision-support system to design efficient interventions, in order to fight against the current pandemic or similar worldwide epidemics that may occur in the future.\\
The rest of the paper is organized as follows. Section~\ref{sec:model} describes the proposed framework, the characteristics of the agents, their set of actions, and the rules to generate the communities in which they are organized. The infection dynamics and the virus transmission mechanisms are explained in detail in the same section. The simulation assumptions, the investigated scenarios, the data collection process, the model’s calibration and the results achieved for a given geography, represented by the State of Delaware, are illustrated in Section~\ref{sec:results}. The research findings are discussed in Section~\ref{sec:dis}, while Section~\ref{sec:conclusion} provides some conclusions and opportunities for future studies.
\section{CoV-ABM: the simulation model}
\label{sec:model}
\begin{figure}
  \centering
  	\includegraphics[totalheight=12cm]{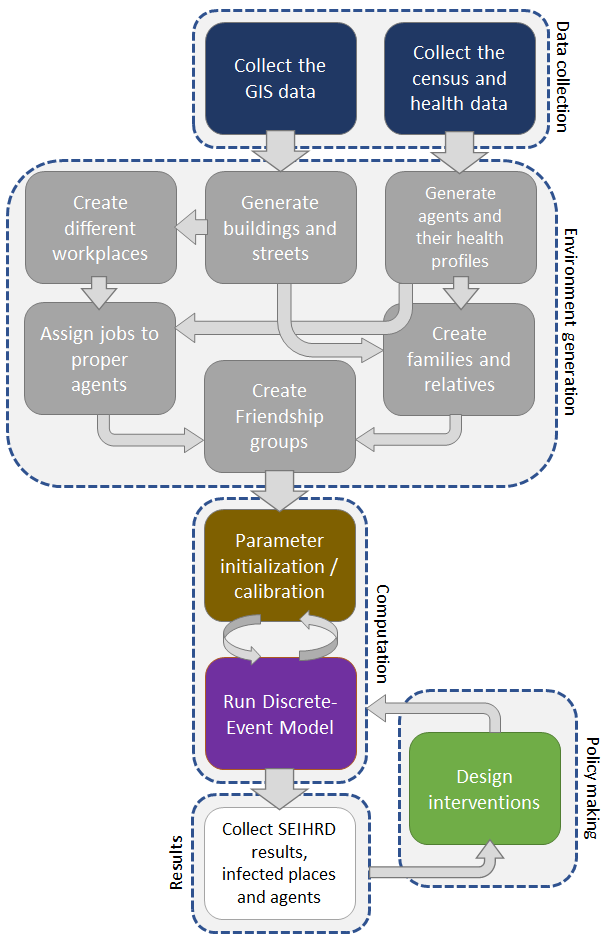}
  \caption{The proposed CoV-ABM framework}
  \label{fig:fig1}
\end{figure}

The paper proposes a novel Agent-Based Model, called CoV-ABM, as a simulator of the SARS-CoV2 infectious disease spread. An overview of the proposed framework is depicted in Figure~\ref{fig:fig1}. Once the corresponding census and health data are collected, the simulator first resorts to the Open Street Map Networkx (OSMnx) library \cite{Boeing2017} for retrieving the GIS data of the geographical areas under investigation. Then, it proceeds with the generation of the buildings, the street maps, and the agents. Based on the health and census data, each agent is associated to a health profile, the households are created and placed in vacant houses. Different workplaces are generated based on the buildings’ information and census data. Agents are then assigned to professional tasks or scholar activities inside these workplaces, based on their profiles and geographical locations. Afterwards, CoV-ABM proceeds with making social communities as described in Section~\ref{sec:com}. Each of the aforementioned steps is described in detail in the following subsections.\\
\subsection{Generating the agents}
\label{sec:agents}
Based on census data, CoV-ABM stochastically generates agents representing citizens living, working, pursuing their needs and travelling inside a geographical environment. To rely on a detailed and realistic simulator, each agent is described in terms of a set of variables, defining its condition at each timestamp. These variables are briefly depicted in Table~\ref{tab:t1}. It should be emphasized that, to the best of our knowledge, this is the first time that such a detailed agent profile is proposed in the literature.\\
\begin{table}
 \caption{Variables defining each agent}
  \centering
  \begin{tabular}{m{3cm} m{1.5cm} m{1.5cm} m{6.5cm}}
    \toprule
    \textit{Variable} & \textit{Notation} & \textit{Type} & \textit{Description} \\
    \midrule
    Name & $u$ & string & A 4-digit string as the agent’s first name\\
    Family  & $v$  & string  & An 8-digit string as the agent’s family ID\\
    Home & $m$  & string  & A GIS location + 3-digit string indictating the building unit\\
    & & & where the agent dwells\\
    Age  & $\zeta$  & integer  & An integer between 1 land 100\\
    Health vulnerability  & $h$  & float  & A real number between 0 and 0.9999\\
    Risk factor  & $\varphi$  & array  & A vector of binary values with respect to different risk factors 1: if the agent has the risk factor, 0: otherwise\\
     Disability & $d$ & binary & 1: severe disability, 0: without disability
\\
     SEIHRD & $s$ & string & The epidemiological state of the agent\\
     Job & $j$ & string & The agent’s job name\\
     Workplace & $w$ & string & A GIS location + 3-digit string to indicate the building unit\\
     & & & where the agent works\\
     Working table & $W^{d,h}$ & array & A matrix $W=(w_{d,h}) \in \mathbb{N}^{7\times 24}$\\
     Location & $\varrho$ & string & The location where the agent is at time $t$\\
     Free & $\phi$ & binary & 1: if the agent is free to perform non-emergency tasks at time, 0: otherwise\\
    \bottomrule
  \end{tabular}
  \label{tab:t1}
\end{table}
When the agents are created, the model associates one with each other to generate families based on the household size, and set their arrangement so to be proportionate to the census data. As an example, if a family includes a kid there must be at least an adult in the same household. In what follows, each agent will be indicated as $a^{u,v}$, whereas $f_{i}=\lbrace a^{v,u}\vert v=i \rbrace$ will represent the ith family. To reduce unnecessary complexity, the agents do not have a gender. Moreover, their disability values, defined in Table~\ref{tab:t1}, are stochastically chosen based on a Bernoulli distribution as follows:\\
\begin{equation}
a^{v,u}(d) \sim Ber(p^{g}_{d})
\end{equation}
where $p^{g}_{d}$ is the percentage of people with severe disabilities within the age-group g, based on the census data. These people seldom do outdoor daily activities, like going for grocery shopping, and mostly rely on someone else for assistance.\\
CoV-ABM defines the “health vulnerability” variable for each agent, which determines how vulnerable or strong an agent can persist against the SARS-CoV2. A strong agent has a higher probability to get recovered after inhaling the virus. Despite some health factors seem to play a key role in the pathogenesis of COVID-19 infection, their relative importance is however still unclear \cite{Jordan2020}. So far, age showed a significant correlation with the fatality rate among infectious people. As of today, more than 70\% of death cases involved individuals older than 65, while the share of fatalities for people under 18 is just 0.04\%. At the same time, recent research findings showed that Body Mass Index (BMI) impacts on the condition of the patients with SARS-CoV2, since obesity compromises pulmonary function and makes the ventilation more difficult. As a consequence, SARS mortality rate significantly increases among patients with BMI values over than 27 \cite{Dietz2020, Kassir2020}. Moreover, Ashokka et al. \cite{Ashokka2020} demonstrated that, although the risk of severe illness for pregnant patients in the COVID-19 pandemic is less than that of similar cases observed during the H1N1 counterpart, the virus seems to exert significant adverse effects in case of childbearing. Indeed, the data provided in their study showed that roughly 9\% of the patients had serious morbidities with respiratory failure or other organs shock. There are also evidences illustrating that the history of smoking is correlated with the deterioration of the patients’ conditions \cite{Vardavas2020}. Finally, recent studies showed that, despite acute respiratory failure is associated with higher mortality, lungs are not the only organs involved since kidney and cardiovascular system health conditions play prominent roles as well \cite{Chu2005, Zheng2020}.\\
In light of these remarks, therefore, we supposed that every agent can have each of these risk factors with a Bernoulli distribution as follows:
\begin{equation}
a^{v,u}(\varphi_{f}) \sim Ber(p^{g}_{f})
\end{equation}
where $p^{g}_{f}$ represents the prevalence of having the factor \textit{f} within the age-group \textit{g}. Moreover, since correlations might exist among these different determinants, we introduced an upper and a lower health bound for each agent, from which the health vulnerability value is stochastically chosen by using a uniform distribution:
\begin{equation}
b_{upper}(a^{v,u}) = min((1-q_{f})^{a^{v,u}(\varphi_{f})} \vert f \in risk factors)
\end{equation}
\begin{equation}
b_{lower}(a^{v,u}) = \prod\limits_{f \in risk factors} (1-q_{f})^{a^{v,u}(\varphi_{f})}
\end{equation}
\begin{equation}
a^{v,u}(h) \sim unif(b_{lower}(a^{v,u}),b_{upper}(a^{v,u}))
\end{equation}
where $q_{f}$ represents the percentage of severe morbidity cases with the risk factor \textit{f} among SARS-CoV2 patients in the age-group \textit{g}. People with health vulnerability values lower than a specific threshold were considered as in need of hospital care and treatment. Since they barely perform shopping or similar tasks, we considered them like agents with a serious disability, i.e. $a^{v,u}(d)=1$. Notice that, in an unrealistic situation where no correlations exist, so that risk factors can be deemed independent one from each other, based on the Bayes’ theorem the probability of their coincidences is the product of the probability of their solely occurrences, which is equal to the lower health bound. At the other extreme, if all the events’ probability spaces are subsets of each other, the probability of their coincidences is given by the probability of the rarest event, which in its turn is equal to the upper health bound. Table~\ref{tab:t2} shows the $p^{g}_{f}$ and $q_{f}$ values for the group composed of people between 25 and 29 years old for a city in the US, based on census data.\\
\begin{table}
 \caption{Risk factors for the age-group 25-29 in the US}
  \centering
  \begin{tabular}{llllll}
    \toprule
    \multirow{2}{*}{\textit{Risk factors}} & \multicolumn{2}{c}{Census-based prevalence $p^{g}_{f}$} & \multicolumn{2}{c}{Severe morbidity probability $q_{f}$}\\
    {} & \textit{Value} & \textit{References} & \textit{Value} & \textit{References} \\
    \midrule
    Age & 1 & - & 0.02 & \cite{Worldometers2020}\\
    Obesity & 0.35 & \cite{Hales2017} & 0.1 & \cite{Kassir2020,Dietz2020}\\
	Pregnancy & 0.08 & Statista\footnote{\url{https://www.statista.com/statistics/295731/us-pregnancy-rates-by-age/}} & 0.09 & \cite{Ashokka2020}\\
	Smoking & 0.16 & \cite{Creamer2019}, CDC\footnote{\url{https://www.cdc.gov/tobacco/data_statistics/fact_sheets/adult_data/cig_smoking/index.htm}} & 0.15 & \cite{Vardavas2020}\\
	Chronic respiratory diseases & 0.001 & \cite{Xie2020} & 0.08 & \cite{Worldometers2020}\\
	Acute renal disease & 0.001 & \cite{USDepartmentofHealthandHumanServices2019} & 0.91 & \cite{Cheng2020, Chu2005}\\
	Heart disease & 0.05 & \cite{USNationalCenterforHealthStatistics2018} & 0.13 & \cite{Worldometers2020}\\
	Diabetes & 0.03 &  \cite{USDepartmentofHealthandHumanServices2020} & 0.09 & \cite{Worldometers2020}\\

    \bottomrule
  \end{tabular}
  \label{tab:t2}
\end{table}
In the CoV-ABM framework two types of agents are generated: those who have a job, in the form of a working or a school activity, which has been assigned to them with a random job-filler algorithm which considers their profiles, and those who aren’t associated to any job. This second group includes job-seekers, whose population size has been set according to the most recent observed unemployment rate, children at preschool ($a^{v,u}(\zeta)<7$) or people older than $x_{retired}$, where this last threshold is given by a stochastic number having a normal distribution as follows:
\begin{equation}
x_{retired} \sim N(\mu_{retired},\sigma_{retired}^{2})
\end{equation}
where   and  represent, respectively, the average and the standard deviation of the retirement age recorded in the geographical area under investigation.\\
Finally, at each timestamp every working agent decides whether to travel to the workplace or not. To rely on a realistic simulator, agents are not always on-time and they might be up to two timestamps late or one timestamp in advance when they go to work or come back home.
\subsection{Generating communities}
\label{sec:com}
In the proposed model, agents belong to different friendship groups or communities of relatives. Friendship groups are composed on average by four individuals, whose membership is not exclusive. This means that agents $a^{v_{1},u_{1}}$ and $a^{v_{2},u_{2}}$ can be friends in more than one group and can share different sets of common friends. At every timestamp, each friendship group stochastically searches among its members with the aim of finding those willing to organize a meetup, where the probability of this event differs according to the specific day in the week and time in the day. The likelihood of successfully arranging a meeting, for example, is much higher for a weekend evening compared to any other evening during the week. If such a meetup is made possible, the available members decide where to go and the place is chosen randomly either from the members’ homes or from open public places in their city, e.g. parks, cafes, shopping centres or stadiums. The model assumes that if an agent attends a meetup it is engaged for a given time interval  and, therefore, it is not able to take part to any other meeting for that period. In what follows, $W^{t}$ will denote the set of willing members at time \textit{$t$}, and \textit{$y$}, the place where they choose to go from timestamp \textit{$t+1$} to \textit{$t+l$}:
\begin{equation}
\forall(v,u) \in W^{t}: a^{v,u}(\phi^{i} \vert i \in \lbrace t+1, \ldots, t+l \rbrace)=0
\end{equation}
\begin{equation}
\forall(v,u) \in W^{t}: a^{v,u}(\varrho^{i} \vert i \in \lbrace t+1, \ldots, t+l \rbrace)=y
\end{equation}
It is worth mentioning that the friendship groups comply with the following rules:\\
\begin{enumerate}
\item Let $a^{v,u}$ denote the first agent included in a group. The other members are selected so that they are approximately at its own age. Specifically, the age of the other agents is taken from the range $a^{v,u}(\zeta)\mp(2\times\sigma_{age})$, where $\sigma_{age}\propto a^{v,u}(\zeta)$. As a consequence, in our framework adolescents or young people have only other adolescents or young people as friends. For older individuals friends are chosen similarly but considering a wider range for the age.
\item Agents younger than a certain age do not join any friendship group. The rationale behind this rule is that children are supposed to meet their friends only at school or preschool and aren’t able to organize meetups in public places.
\end{enumerate}
Beside friendship groups, the model includes also communities of relatives involving different families. In this case the membership is exclusive so that each family belongs to at most one of these groups. At every timestamp, each community stochastically searches among its families to find those available to organize a meetup. A family is declared available when at least half of its members are willing to attend the meeting. Once a meeting is set up and its location is chosen, the available family members will attend it and won’t be free for a certain time.\\
Notice that, unlike other agent-based models proposed in the literature which resort to random graphs to represent interactions among the agents, our model relies on a more detailed relationship framework to better reflect the real-world interpersonal interactions. This feature plays an important role since it allows to study quarantine scenarios and social distancing impositions with higher realism and precision. For instance, if people are forced to forego the meetups with their friends due to lockdown or other social restrictions, they are still in contact with their own family members, they might still visit some of their relatives, like grandparents, or they might want to limit the meeting locations to their own houses. As far as we know, the models so far proposed lack such representation capabilities.
\subsection{Agents’ needs and actions}
\label{sec:needs}
As the discrete-event model starts, each family is stochastically associated to a variety of needs which have to be fulfilled. These needs can take the form of \textit{emergency needs}, which are strictly connected to the family members’ health conditions, and of \textit{daily needs}, which are instead referred to every-day activities such as food provisioning, shopping, refueling and so on. The probability of having daily needs depends on the frequency of each need in the real society and it is here assumed to be equal for all families. The likelihood for an emergency need to arise is, instead, dependent on each member’s health conditions, $a^{v,u}(h)$ and $a^{v,u}(s)$, as follows:
\begin{equation}
v_{emergency}^{regular}(u)^{t} \sim Ber(p_{emergency}^{regular}a^{v,u}(h))
\end{equation}
\begin{equation}
v_{emergency}^{COVID}(u)^{t} \sim Ber((p_{i\rightarrow h}^{t})^{v,u}); \lbrace (v,u)  \vert a^{v,u}(s) = infected\rbrace
\end{equation}
where $p_{emergency}^{regular}$ is the likelihood of being hospitalized for a “highly strong agent”, defined in this study as an agent with a health vulnerability value near to 1, as indicated in Table~\ref{tab:t1}. Whereas $p_{i\rightarrow h}^{t}$ represents the
likelihood of being hospitalized due to COVID-19. This second term will be explained in more details in Section~\ref{sec:seihr}.\\
Emergency needs involve a given family member and can be satisfied at a specific hospital located in the county where he/she lives. When an emergency need arises, the affected agent immediately plans to go to the hospital with the aim of fulfilling that need. In case of disability or precarious state, a good health adult belonging to the same family is randomly selected to help the member in need.\\
In our model, the length of staying (LOS) at the hospital for patients affected by COVID-19 is defined as the number of days elapsed between hospitalization and discharge (in case of both recovery and death), while for the other patients it is set to a stochastic number. Notice that, the average LOS value may vary from city to city and for different diseases. However, for the sake of simplicity its value was here set to 4.5 days, which corresponds to the average hospital LOS recorded in the US \cite{Witt2006}.\\
Harini et al. observed that non-transformed distributions cannot fit the LOS data appropriately, and suggested to use the gamma-pareto distribution for hospitals LOS fitting purposes \cite{Harini2018}. In our model, therefore, each non-COVID patient stays at the hospital for a period of  days, which follows a gamma-pareto distribution \cite{Alzaatreh2012}\\
\begin{equation}
L \sim Gamma-Pareto(\alpha_{LOS},c_{LOS},\theta_{LOS})
\end{equation}
with parameter values $\alpha_{LOS}=7$, $c_{LOS}=0.2$ and $\theta_{LOS}=1$, as suggested in \cite{Harini2018}.\\
Daily needs are by their nature less urgent and can ideally be satisfied by any eligible member of the family. The eligibility is given to not-disabled members showing health vulnerability values greater than a given threshold. In particular, as soon as an eligible member manages to find the time and the right place is open, he/she goes there to fulfill the need. In the event that several places are available, the priority is given to the closer destination. To this aim, we resorted to a graph-based shortest path algorithm proposed in NetworkX \cite{Hagberg2008}. Furthermore, to govern the complexity, we assumed that on average each family daily need requires one hour to be satisfied.\\
The model uses a Bernoulli trial to generate new needs for each family, the probability of which is given by:
\begin{equation}
v_{daily}^{n , t} \sim Ber(\vartheta_{n}dt/24) ; n \in N
\end{equation}
where $\vartheta_{n}$ is the frequency of expressing the daily need $n$. $dt$ denotes the timestep and $N$ is the set of different daily needs.\\
When an agent is selected to satisfy a need, either an emergency or a daily one, it won’t be free to accomplish other tasks until the need is fulfilled. For a family, if a daily need is not yet satisfied and a new similar need is generated, these two needs are merged into a single one.
\subsection{Infection dynamics}
\label{sec:seihr}
In this paper we propose a novel way of disease transmission among the agents. Specifically, we assume that the individual suffering from the disease can infect other agents indirectly by spreading the virus in the environment. As of today, it is believed that COVID-19 is often transmitted through respiratory droplets produced by coughing and sneezing. This virus can however remain suspended in the air for a significant period of time \cite{Bar-On2020} and may also persist on surfaces like metal, plastic or glass for up to 9 days \cite{Kampf2020}. Developing a model which accounts for such an important characteristic is fundamental to properly represent the spread of SARS-CoV2. Hence, unlike other agent-based models which assume directly contacts as the only mean of transmission, we suppose that the environment itself (the air and/or a surface) can convey the virus from one agent to the other.\\
In the CoV-ABM framework each location, either indoor or in the form of an outdoor area, can contain a given amount of virus. The volume of the virus persisting in the given location divided by the volume of the location itself determines the probability of infection. The persistence of the virus is not yet exactly known and is estimated to range between two hours and several days, depending on the environmental conditions \cite{Kampf2020}. However, we can suppose that in a given venue COVID-19 gets inactivated in the environment after $k$ hours on average. Thus, after each timestamp a portion of virus equal to $\delta=dt/k$ at each location disappears. Note that the value of $k$ can change by applying different mitigation measures, such as the use of detergents and UV lamps. According to this model, an agent is be able to transmit the virus to any other agent residing or visiting the locations it stayed or passed by. To the best of our knowledge, none of the existent agent-based models in the literature relies on such an assumption.\\
\begin{figure}
  \centering
  	\includegraphics[totalheight=5cm]{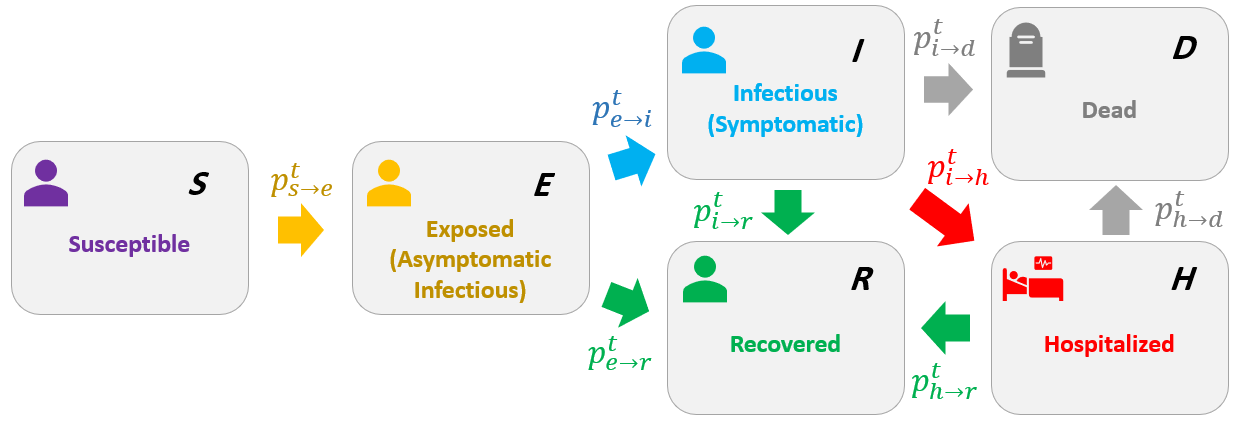}
  \caption{Schematic diagram of the proposed SEIHRD model, where arrows indicate possible transitions with the corresponding probability notations}
  \label{fig:fig2}
\end{figure}
In this study an extension of the SEIR model, called SEIHRD, is used to shape the epidemiological phases of the disease. As discussed earlier in Section~\ref{sec:s1}., some SEIR extensions have been proposed in the literature addressing the spread of COVID-19 \cite{Ivorra2020, Yang2020} and \cite{Kantner2020}. As it is shown in Figure~\ref{fig:fig2}, our SEIHRD model comprises 6 states: Susceptible (S), Exposed or Asymptomatic Infectious (E), (Symptomatic) Infectious (I), Hospitalized (H), Recovered (R) and Dead (D). For the sake of simplicity, unlike some recent works \cite{Moghadas2020}, we do not separate ICU beds and regular beds.\\
By default, all the agents are susceptible except for the initial spreaders, called seeds:
\begin{equation}
a^{v,u}(s^{t=0})=\begin{cases}
               susceptible ; \forall (v,u) \notin seeds\\
               exposed ; \forall (v,u) \in seeds
            \end{cases}
\end{equation}
At each timestamp, exposed and infectious agents spread the virus in the locations where they resided or passed by.  For each location $y$, the infection is calculated as:
\begin{equation}
y(\Delta^{t})=y(\Delta^{t-1})(1-\delta)+\sum\limits_{(v,u)\in A_{y}^{t}} \epsilon^{a^{v,u}(s^{t})} ; A_{y}^{t}=\lbrace (v,u) \vert a^{v,u}(\varrho^{t})=y \rbrace
\end{equation}
where $y(\Delta^{t})$ is the volume of active viruses inside $y$, and $\epsilon^{a^{v,u}(s^{t})}$ represents the average quantity of virus being spread in the environment, at a given timestamp, by an agent who is in the epidemiological state $s^{t}$. $A_{y}^{t}$ is the set of agents located in $y$ at time $t$.\\
The probability of being infected for any susceptible agent inside the given area is then determined by a Bernoulli trial with transmission probability $p_{s \rightarrow e}^{t}$. Specifically, for agent $a^{v,u}$ at time $t$ this probability can be computed as:
\begin{equation}
p_{s \rightarrow e}^{t} = \alpha(a^{v,u}(\varrho^{t})(\Delta^{t})/a^{v,u}(\varrho^{t})(\Omega^{t}))
\end{equation}
where $\alpha$ is the infectiousness rate of the disease, which is a real number between 0 and 1, and $y(\Omega^{t})$ denotes the volume of location $y$.\\
If the agent receives the virus, during the “exposed” phase it won’t show any clinical symptom. Therefore, it will work and live like a susceptible agent while spreading the virus over the places it visits or resides in. After a stochastic period, whose average is $L_{e}$(hours), the agent is either \textit{“infectious”} or \textit{“recovered”}:
\begin{equation}
p_{e \rightarrow i}^{t} = \beta(1-\rho_{er}(a^{v,u}(h^{t})))
\end{equation}
\begin{equation}
p_{e \rightarrow r}^{t} = \beta \rho_{er}(a^{v,u}(h^{t}))
\end{equation}
where $\beta=dt/L$, and $\rho_{er}$ is the rate of being directly recovered after being asymptomatic infectious for a highly strong agent. Thus, as $a^{v,u}(h^{t})$ decreases, the probability of being “infectious” raises.
During the “infected” phase the agent shows symptoms; it still spreads the virus but with a different rate $\epsilon^{infectious}$. The average time length of being infectious for a highly strong agent is given by $L_{i}$(hours). 
Let $\gamma=dt/L_{i}$ be the rate of being “recovered” for a highly strong agent. The probability of becoming “hospitalized”, “recovered” or “dead” at each timestamp is:
\begin{equation}
p_{i \rightarrow r}^{t} = \gamma (a^{v,u}(h^{t}))
\end{equation}
\begin{equation}
p_{i \rightarrow h}^{t} = \rho_{ih} (a^{v,u}(h^{t}))
\end{equation}
\begin{equation}
p_{i \rightarrow d}^{t} = \rho_{id}(1-\gamma(a^{v,u}(h^{t})))
\end{equation}
where $\rho_{ih}$ and $\rho_{id}$ represent the hospitalization and the death rate, respectively, among the strong individuals who are not hospitalized. For hospitalized patients, instead, the probability of being “recovered” is still the same, since there is no evidence illustrating that hospitalized people have shorter period of infection: $p_{i \rightarrow r}^{t}=p_{h \rightarrow r}^{t}$ . However, due to the presence of dedicated facilities such as Intensive Care Units and of special medical equipment, hospitalization reduces the death rate. Consequently, we propose to compute this probability as:
\begin{equation}
p_{h \rightarrow d}^{t} = p_{i \rightarrow d}^{t}/H_{q}^{t}
\end{equation}
where $H_{q}^{t}$ represents the efficiency of hospital $q$ at time $t$, called “hospital factor”. If hospital $q$ has a treatment capacity equal to $C_{q}^{t}$, the hospital reaches its maximum efficiency if there are less than $C_{q}^{t}$ agents being hospitalized. If the number of hospitalized individuals at $q$ increases, the hospital factor plunges due to the saturation of the structures and the scarcity of available care beds. Let $n^{t}$ be the number of patients in hospital $q$ at time $t$; we propose a modified pareto curve to calculate the efficiency values as follows:
\begin{equation}
H_{q}^{t} = 1+\tau_{0}C_{q}^{t}/(C_{q}^{t}+\tau_{1}n^{t})
\end{equation}
\begin{figure}
  \centering
  	\includegraphics[totalheight=5cm]{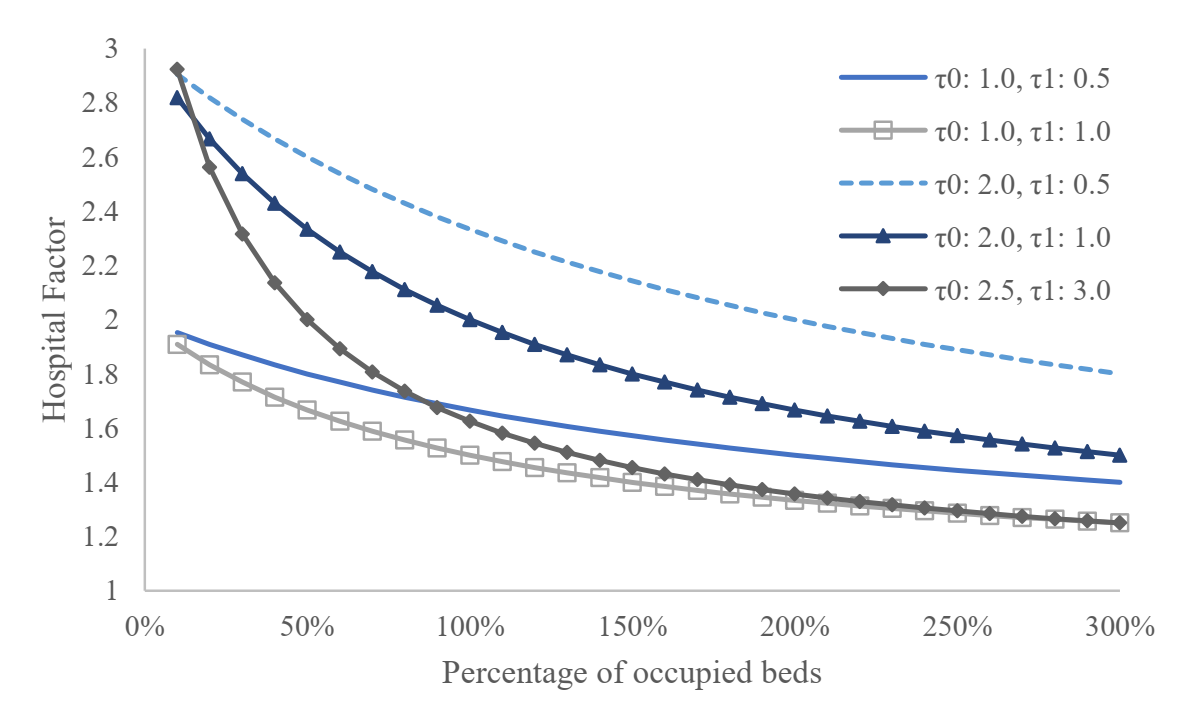}
  \caption{Hospital factor with different parameter values}
  \label{fig:fig3}
\end{figure}
Figure~\ref{fig:fig3} illustrates the hospital factors with different parameter values. These values may change from city to city and hospital to hospital, so a parameter tuning needs to be accomplished. The pseudocodes of the proposed SEIHRD diffusion model and the main discrete-event of CoV-ABM are provided in Algorithm~\ref{alg:seir} and Algorithm~\ref{alg:covabm} (see appendix~\ref{app:alg}).
\section{Simulation results}
\label{sec:results}
\begin{figure}
  \centering
  	\includegraphics[totalheight=9cm]{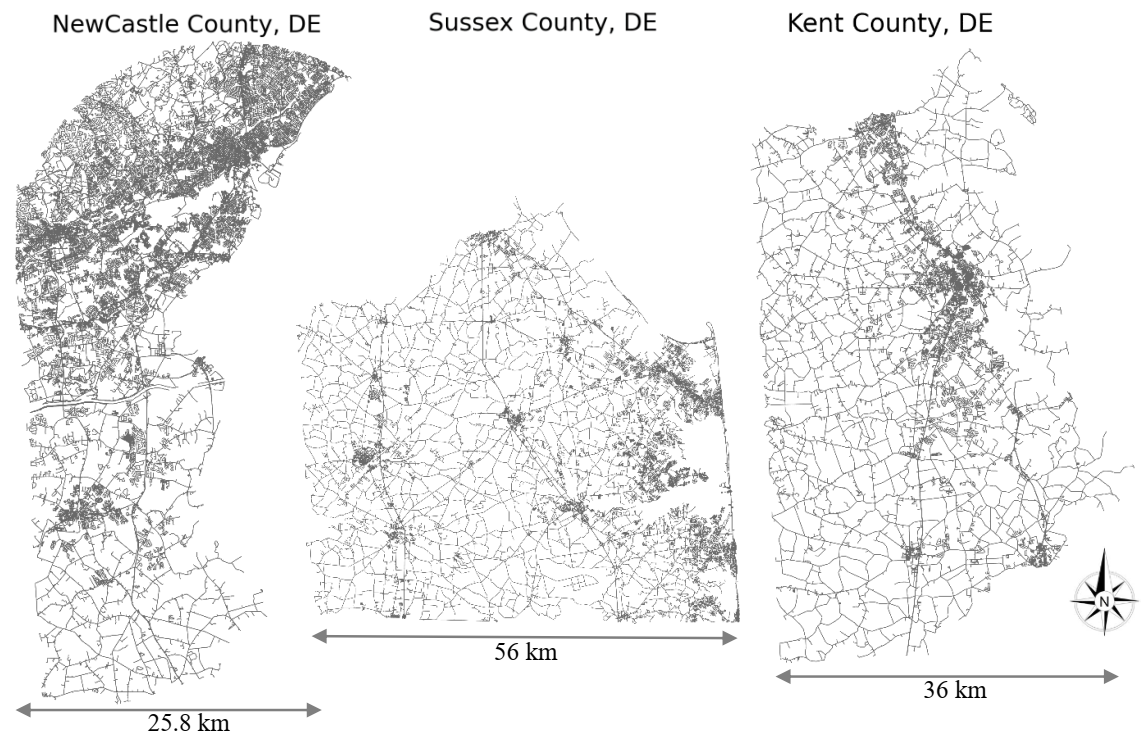}
  \caption{Street networks of New Castle County (left), Sussex County (middle) and Kent County (right)}
  \label{fig:fig4}
\end{figure}

In this section we illustrate how the model can simulate the spread of COVID-19 in the State of Delaware, USA, which has a population around one million people. This state is divided into three counties, New Castle, Sussex, and Kent, with populations over 558, 234 and 180 thousand, respectively. Delaware is one of the smallest states in the US and ranked 46th out of the fifty US states in terms of population. However, the virus there has been spreading widely in the recent weeks. As of today, Delaware is ranked 6th in the US in terms of total COVID-19 positive cases per one million population, with near 10,000 approved cases \cite{USDepartmentofHealthandHumanServices2020}. Figure~\ref{fig:fig4} depicts the network of streets of the three counties, separately.\\
\subsection{Simulation assumptions and different scenarios}
\label{sec:assumptions}
In order to govern the complexity of the simulator we assumed the environment to be isolated. Consequently, the model doesn’t account for incoming or outgoing travels from the region at hand. Notice that, this hypothesis well fits with our simulation environment since the State of Delaware occupies the north-eastern portion of the Delmarva Peninsula, besides some islands and territory within the Delaware River, and is therefore less connected to other states compared to inland states.\\
Further assumptions were then introduced to cope with the lack of complete information, preserving however the compatibility and the adherence to detailed census data. From one side, some buildings in the state were randomly assigned to different uses, being unaware of their actual usage. From the other, the unavailability of census data reporting an exhaustive list of job titles and workplace types, required to synch the environment generator, forced us to simplify jobs, workplaces and the corresponding working staffs. Table~\ref{tab:t3} shows the different working and living spaces investigated in the present study. The amount of working hours was randomly assigned to each worker, by assuming that no healthy individual works less than 6 hours or more than 10 hours, daily. Moreover, some workplaces such as hospitals, public transportations or police stations were supposed to be always up and running.\\
\begin{table}
 \caption{Categories of workplaces and living spaces}
  \centering
  \begin{tabular}{m{2cm} m{6cm}}
    \toprule
    \textit{Categories}  & \textit{Working and living spaces}\\
    \midrule
    Education & University, College, School, Kindergarten\\
	Emergency & Hospitals, Fire stations, Police stations\\
	Amenity & Parks, Cafés, Restaurants, Gyms, Casinos\\
	Event & Pubs, Stadiums\\
	Crucial & Banks, Post Offices, Courts, Supermarkets, Drugstores, Gas stations, Public transportations, Crucial factories, Power plants\\
	Cultural & Museums, Churches, Mosques, Zoos\\
	Private & Shops, Industrial plants, Offices\\
	Residential & Apartments, Detached houses, Villas\\
    \bottomrule
  \end{tabular}
  \label{tab:t3}
\end{table}
In our framework, friends and relatives can get together in one of the places comprised in the categories Residential, Amenity, Event or Cultural, with different probabilities, which take into account the specificity of both location and meeting attendees. For example, the likelihood for a reunion to take place in a café is higher than in a museum; at the same time, friends tend to meet in Stadiums, Cafés, Gyms, Parks, while relatives are more inclined to visit each other at their own houses.\\
According to the hypothesis made in \cite{Cliff2018}, we further assumed that schools and kindergartens admit people from the closer districts. Hence, the school located closest to a child’s home had a higher probability to be selected. 
Finally, to limit the computational effort the model was run on each county separately, and the results were then summed together. Notice that, the same framework can be run on the whole country with few minor modifications at the expense of an increase of the simulation time.\\
With the aim of studying the usefulness of some virus counter-measures and investigate at what extent they can affect the curves of infected people and cause different spread outcomes, two alternative scenarios were defined as follows:
\begin{itemize}
\item No interventions: All places remain open and practicable full-time. The probability of visiting friends is $1/14 (day^{-1})$ and relatives is $1/30 (day^{-1})$.
\item Quarantine: All the places in the Education, Amenity, Cultural and Event categories are closed with their staff staying at home. The probability of visiting friends and relatives drops to $1/30 (day^{-1})$ and $1/60 (day^{-1})$, respectively. In this case, therefore, people still have contacts with friends and relatives; however, meeting destinations are no longer represented by public places like parks or cafés, closed by the restrictions, but are replaced by the individuals’ own houses.
\end{itemize}
It is worthwhile to observe that the previous scenarios were conceived for the purpose of analyzing the effectiveness on the mitigation of the outbreak of two specific types of interventions, represented by the drastically reduction of social meetings and the closure of non-critical places.
\subsection{Simulation of agents on census data}
\label{sec:census}
\begin{figure}
  \centering
  	\includegraphics[totalheight=10cm]{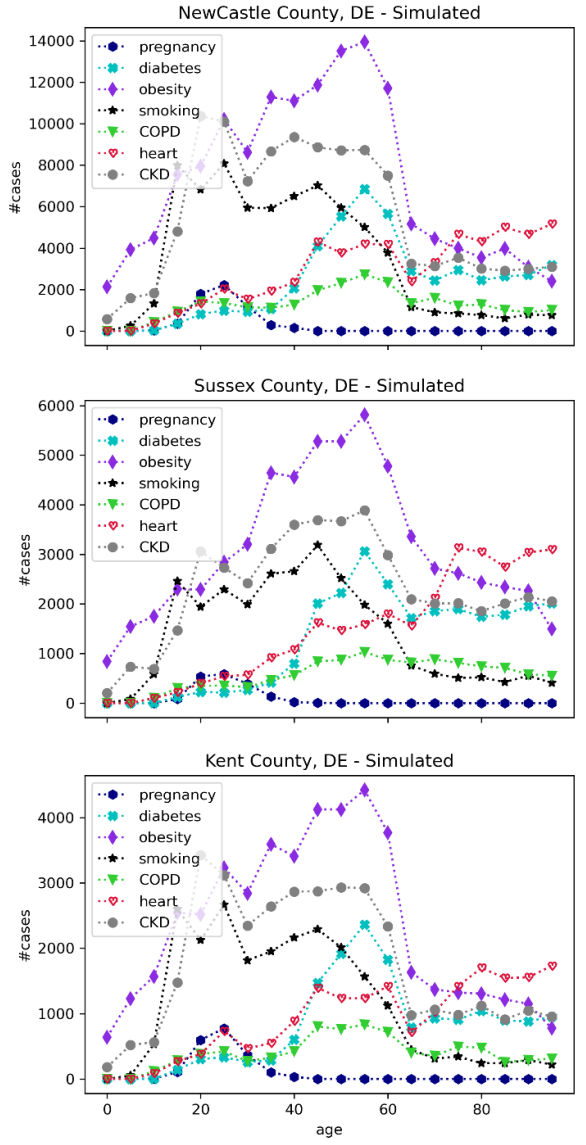}
  \caption{Risk factors among the simulated agents}
  \label{fig:fig5}
\end{figure}

In order to generate the simulated environment and the agents’ profiles all the necessary demographic and health information about Delaware’s counties were gathered from US census and US health reports. In particular, data on each county’s hospitals and relative capacities were collected from the American Hospital Directory website\footnote{\url{https://www.ahd.com/states/hospital_DE.html}}, whereas the Open Street Map tools\footnote{\url{https://wiki.openstreetmap.org/wiki/Main_Page}}, e.g. OSMnx, were used to retrieve information about locations, building usage and street graphs.\\
Once the agents were generated based on the census and health data, their profiles were analysed to verify the validity of the achieved population. Table~\ref{tab:t4} compares the population of the simulated agents, divided into 5 age-groups, with the real-world census, and shows the close correspondence between the created and the actual population’s demography. Figure~\ref{fig:fig5}, instead, illustrates the number of generated agents affected by the different risk factors in each county.\\
\begin{table}
 \caption{Population of agents in different age-groups}
  \centering
  \begin{tabular}{lllllll}
    \toprule
    \multirow{2}{*}{\textit{Risk factors}} & \multicolumn{2}{c}{New Castle County} & \multicolumn{2}{c}{Sussex County} & \multicolumn{2}{c}{Kent County} \\
    {} & \textit{Simulated} & \textit{Real-world} & \textit{Simulated} & \textit{Real-world} & \textit{Simulated} & \textit{Real-world}\\
    \midrule
    < 5 & 5.5\% & 5.6\% & 5.2\% & 5.2\% & 5.5\% & 6.3\%\\
    5 - 17 & 16.3\% & 15.7\% &14.8\% & 14.0\% &16.3\% & 17\%\\
    18 - 44 & 37.1\% &36.1\% & 30.3\% & 27.1\% & 36.9\% & 35.6\%\\
    45 - 64 & 25.3\% & 26.7\% & 25.1\% & 28.4\% & 25.1\% & 25.3\%\\
	65 + & 15.9\% & 15.6\% & 24.6\% & 25\% & 16.2\% & 15.8\%\\
    \bottomrule
  \end{tabular}
  \label{tab:t4}
\end{table}
\subsection{Simulating the spread of SARS-CoV2}
\label{sec:sim}
Simulations were performed by using a mid-range personal laptop with an i7 processor and 8GB of memory on an Ubuntu 18 operating system. The simulator was coded in Python 3.8 and the model was run on each scenario 5 times iteratively, to average the results. The runtime of each iteration for the counties Kent, Sussex and New Castle was roughly 2, 2.5 and 7 hours, respectively.\\
\subsubsection{Simulator calibration}
\label{sec:calibration}
\begin{figure}
  \centering
  	\includegraphics[totalheight=5.5cm]{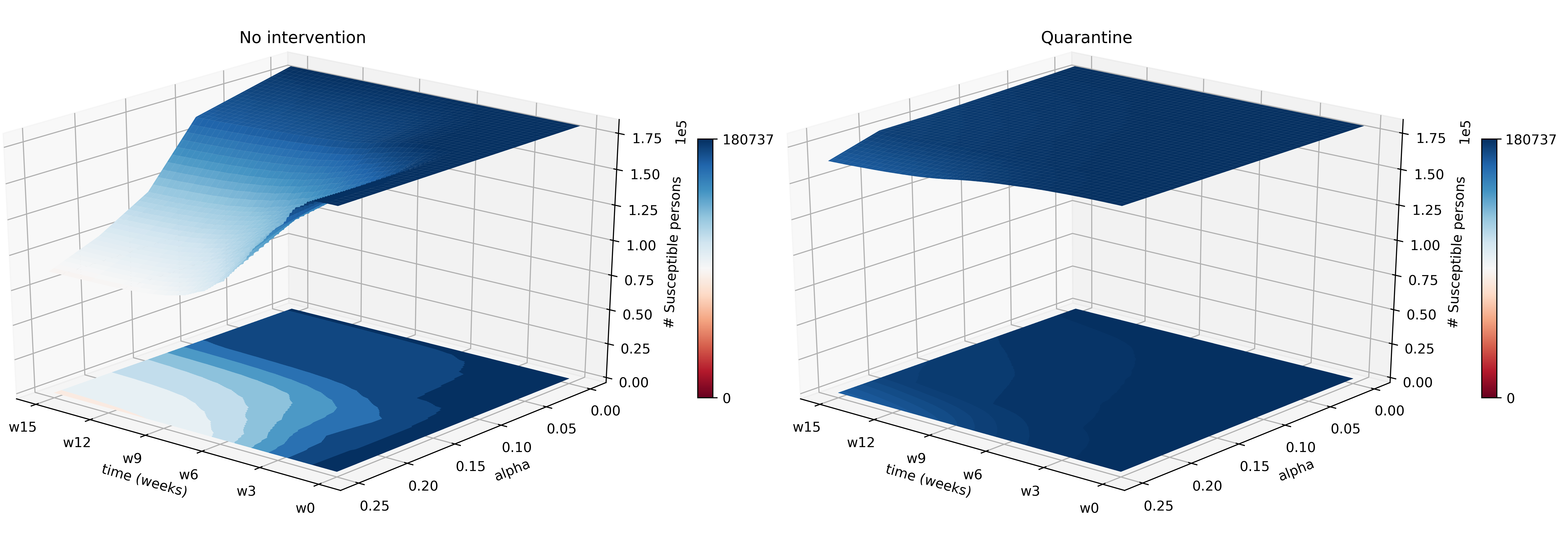}
  \caption{The impact of $\alpha$ on the number of susceptible agents in Kent County with No intervention (left) vs. Quarantine (right)}
  \label{fig:fig6}
\end{figure}

\begin{figure}
  \centering
  	\includegraphics[totalheight=5.5cm]{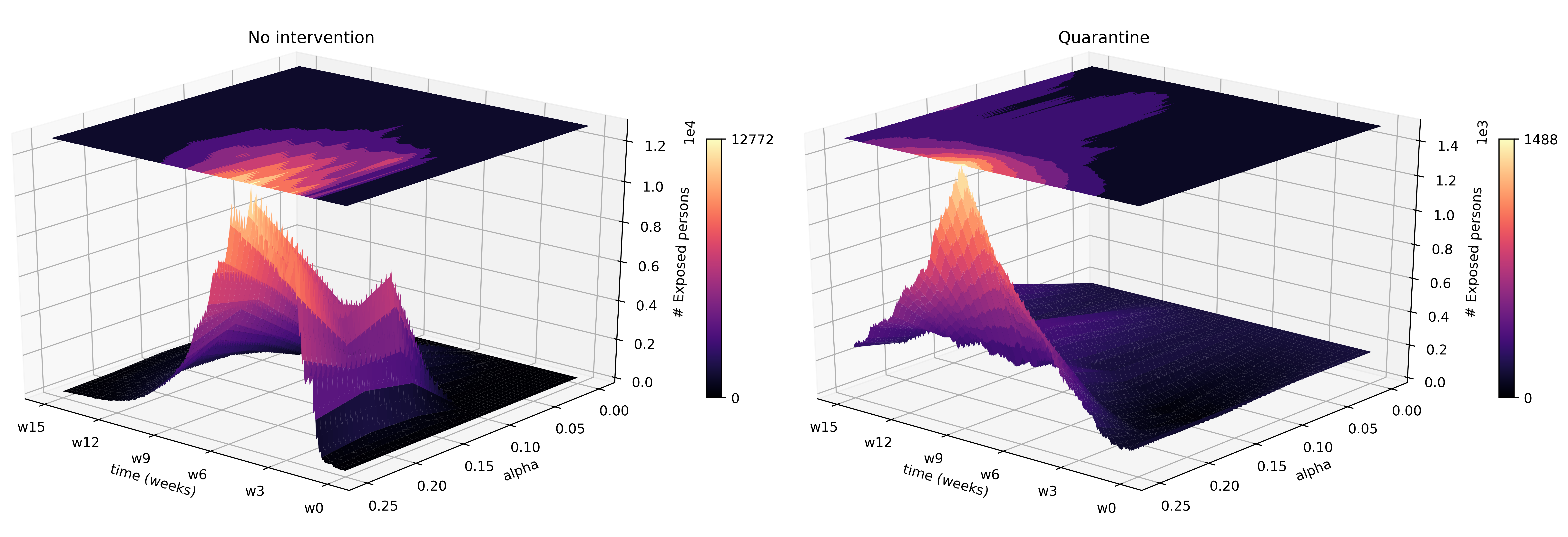}
  \caption{The impact of $\alpha$ on the number of exposed agents in Kent County with No intervention (left) vs. Quarantine (right)}
  \label{fig:fig7}
\end{figure}

\begin{figure}
  \centering
  	\includegraphics[totalheight=5.5cm]{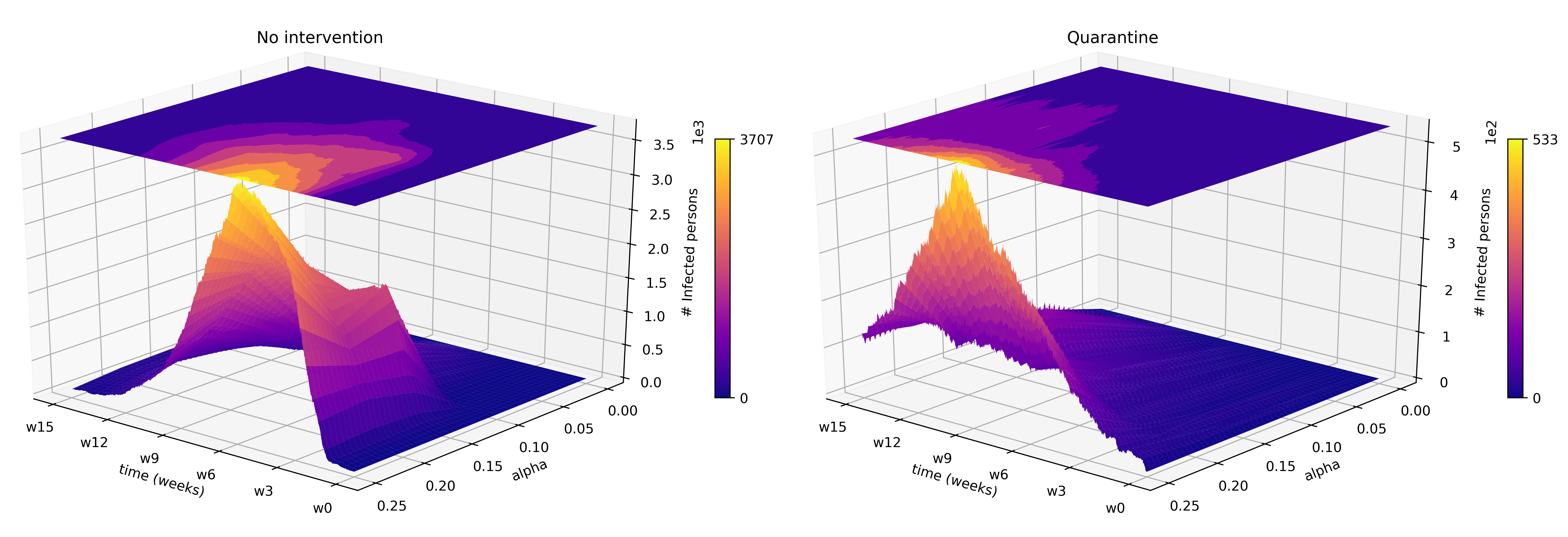}
  \caption{The impact of $\alpha$ on the number of infectious agents in Kent County with No intervention (left) vs. Quarantine (right)}
  \label{fig:fig8}
\end{figure}

\begin{figure}
  \centering
  	\includegraphics[totalheight=5.5cm]{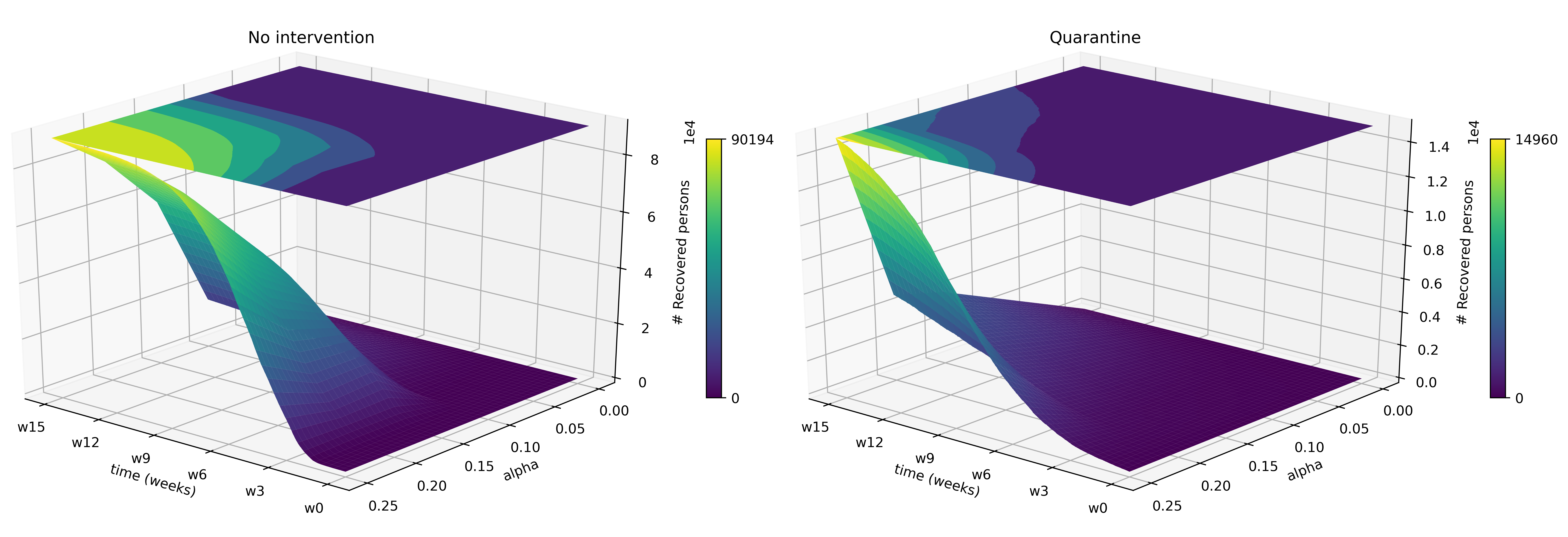}
  \caption{The impact of $\alpha$ on the number of recovered agents in Kent County with No intervention (left) vs. Quarantine (right)}
  \label{fig:fig9}
\end{figure}

\begin{figure}
  \centering
  	\includegraphics[totalheight=5.5cm]{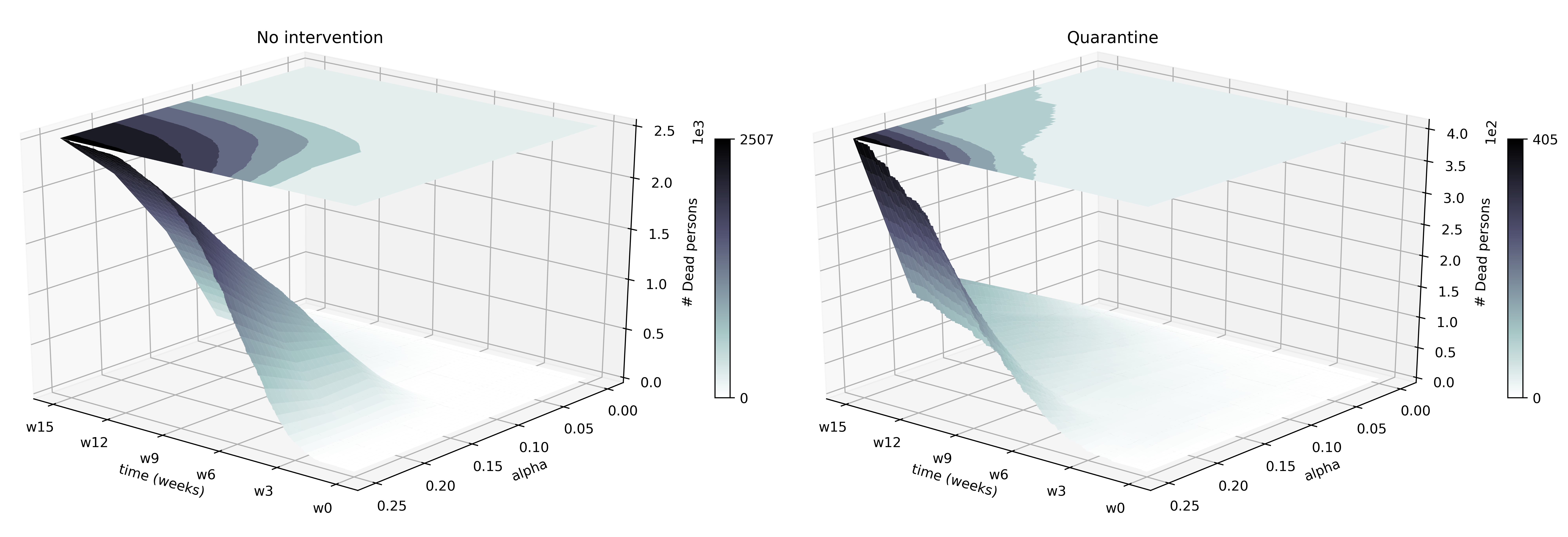}
  \caption{The impact of $\alpha$ on the number of dead agents in Kent County with No intervention (left) vs. Quarantine (right)}
  \label{fig:fig10}
\end{figure}

In order to run the simulator on the State of Delaware, we first set some initial parameters, described in Section~\ref{fig:fig2}., as shown in Table~\ref{tab:t5}.\\
\begin{table}
 \caption{Parameters and their values for simulating on the State of Delaware}
  \centering
  \begin{tabular}{m{2cm} m{6.5cm} m{2cm} m{2cm}}
    \toprule
    \textit{Parameter} & \textit{Description} & \textit{Estimated value} & \textit{References}\\
    \midrule
    $dt$ & Timestamp & 1 (hour) & - \\
    $hr$ &  Average time length of virus persistence & 20 (hours) & \cite{Kampf2020}\\
     & in the environment. &  & \\
$\alpha$ & Infectiousness rate & 0 to 0.25 (1/hour) & Estimated\\
$L_{e}$ & Average time length of being exposed & 120 (hours) & \cite{Li2020,Wang2020}\\
$L_{i}$ & Average time length of being infectious & 110 (hours) & \cite{Shoukat2020}\\
$\rho_{er}$ & Proportion of cases having no symptoms & 80\% & \cite{WorldHealthOrganization2020,Bai2020}\\
$\rho_{ih}$ & Hospitalization rate for an infectious case without risk factors & 10\% & Assumed\\
$\rho_{id}$ & Death rate for an infectious case without risk factors & 5\% & Assumed\\
$\epsilon^{exposed}$  & Average quantity of virus being spread hourly in the environment by an asymptomatic agent & 0.5 (1/hour) & Assumed\\
$\epsilon^{infectious}$  & Average quantity of virus being spread hourly in the environment by a symptomatic agent & 1 (1/hour) & Assumed\\
    \bottomrule
  \end{tabular}
  \label{tab:t5}
\end{table}
As it is defined in Section~\ref{sec:seihr}, the infectiousness rate $\alpha$ determines how easily the virus transmits to a susceptible individual. The exact value of the infectiousness rate is unknown and depends on several factors, such as the virus characteristics, the environment (e.g. humidity, temperature, wind speed) and the personal protective behaviour. As an example, facial masking and careful hand hygiene can reduce the transmission, so that the infectiousness rate in societies with rigorous protective behaviours is usually lower.\\
In order to figure out how the infectiousness rate affects the viral spread dynamics and to properly set its value, we performed several simulations on the Kent county with different  values, ranging from 0 to 0.25 with a step size equal to 0.05. Figures~\ref{fig:fig6} to \ref{fig:fig10} depict the impact of the infectiousness rate on the SEIHRD phases’ surfaces. As it can be seen, high  values result in more infected people and more deaths. Moreover, from the same charts it can be inferred that quarantine and lower infectiousness rates make the whole spread process slower and longer, eventually leading to fewer casualties.\\
\begin{figure}
  \centering
  	\includegraphics[totalheight=8cm]{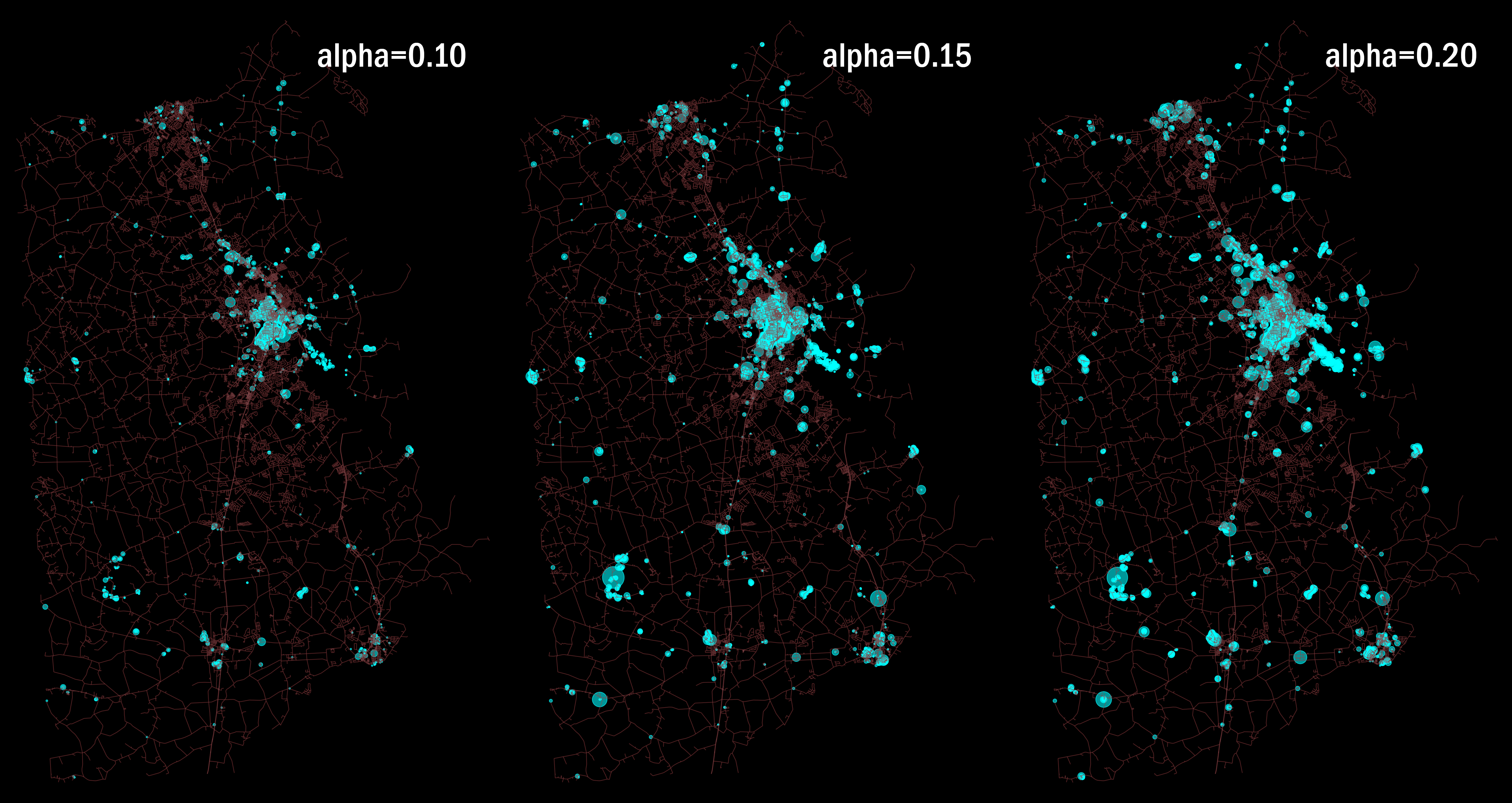}
  \caption{Infected places in Kent County after 7 weeks:$\alpha=0.1$(left), $\alpha=0.15$(middle), $\alpha=0.2$(right)}
  \label{fig:fig11}
\end{figure}
Figure~\ref{fig:fig11} depicts the infected places in Kent County, with different infectiousness rate values, at the corresponding infection peaks. Notice that, due to the virus inactivation period, the highlighted locations only represent the places where the virus still persists and is not disabled. The blue circles are proportional to the virus densities inside the corresponding areas.\\
Once observed the effect of the infectiousness rate at different intensity levels and compared it with the actual curves derived from the available reports, we consequently set it at $\alpha=0.15$ to proceed with the remaining simulations.\\
\subsection{Simulation results for Delaware}
\label{sec:delaware}
\begin{figure}
  \centering
  	\includegraphics[totalheight=8cm]{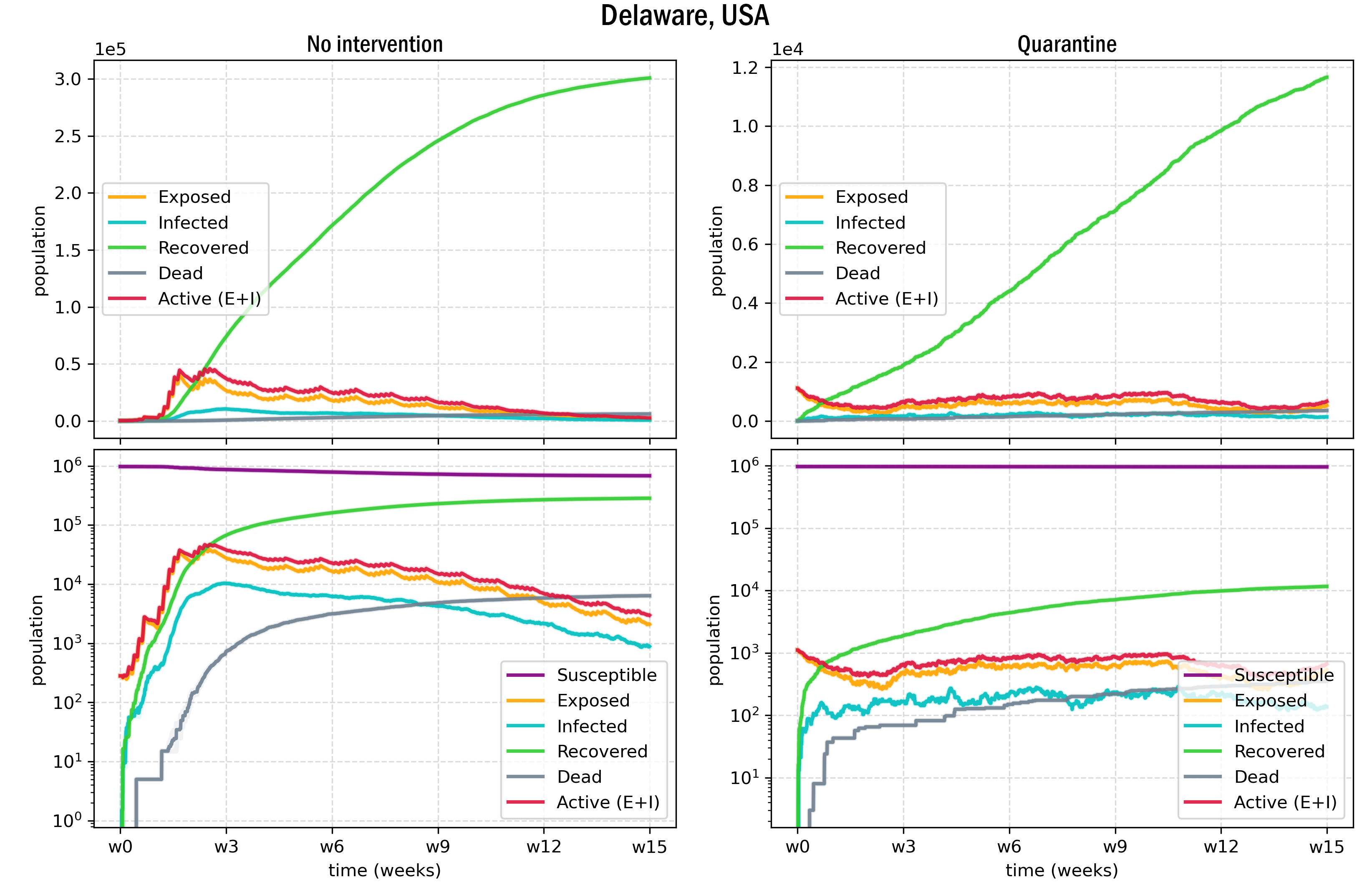}
  \caption{SEIHRD dynamics in the State of Delaware when $\alpha=0.15$ - No intervention (left) vs. Quarantine (right)}
  \label{fig:fig12}
\end{figure}
\begin{figure}
  \centering
  	\includegraphics[totalheight=8cm]{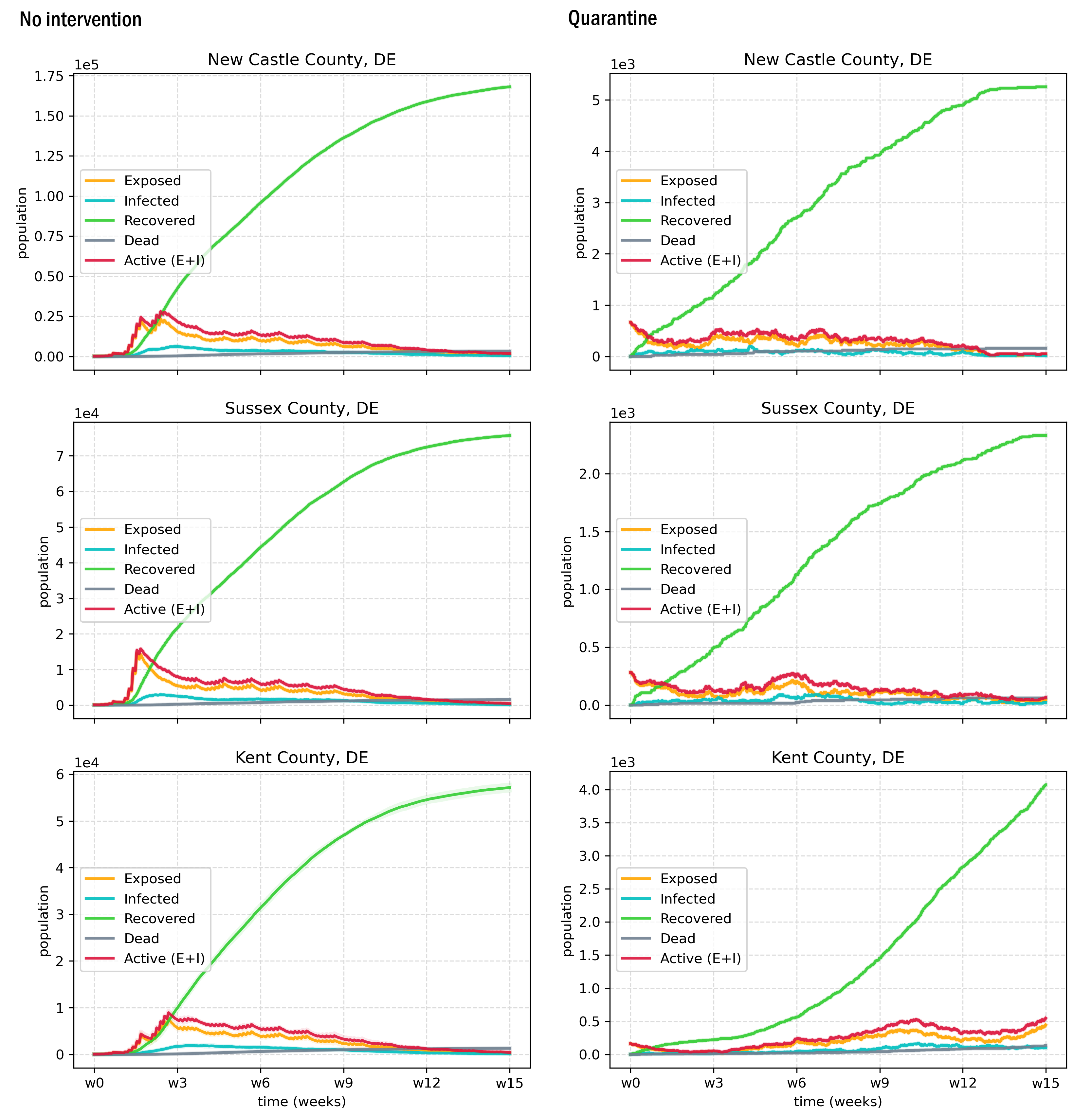}
  \caption{SEIHRD dynamics in each county when $\alpha=0.15$ - No intervention (left) vs. Quarantine (right)}
  \label{fig:fig13}
\end{figure}
\begin{figure}
  \centering
  	\includegraphics[totalheight=8cm]{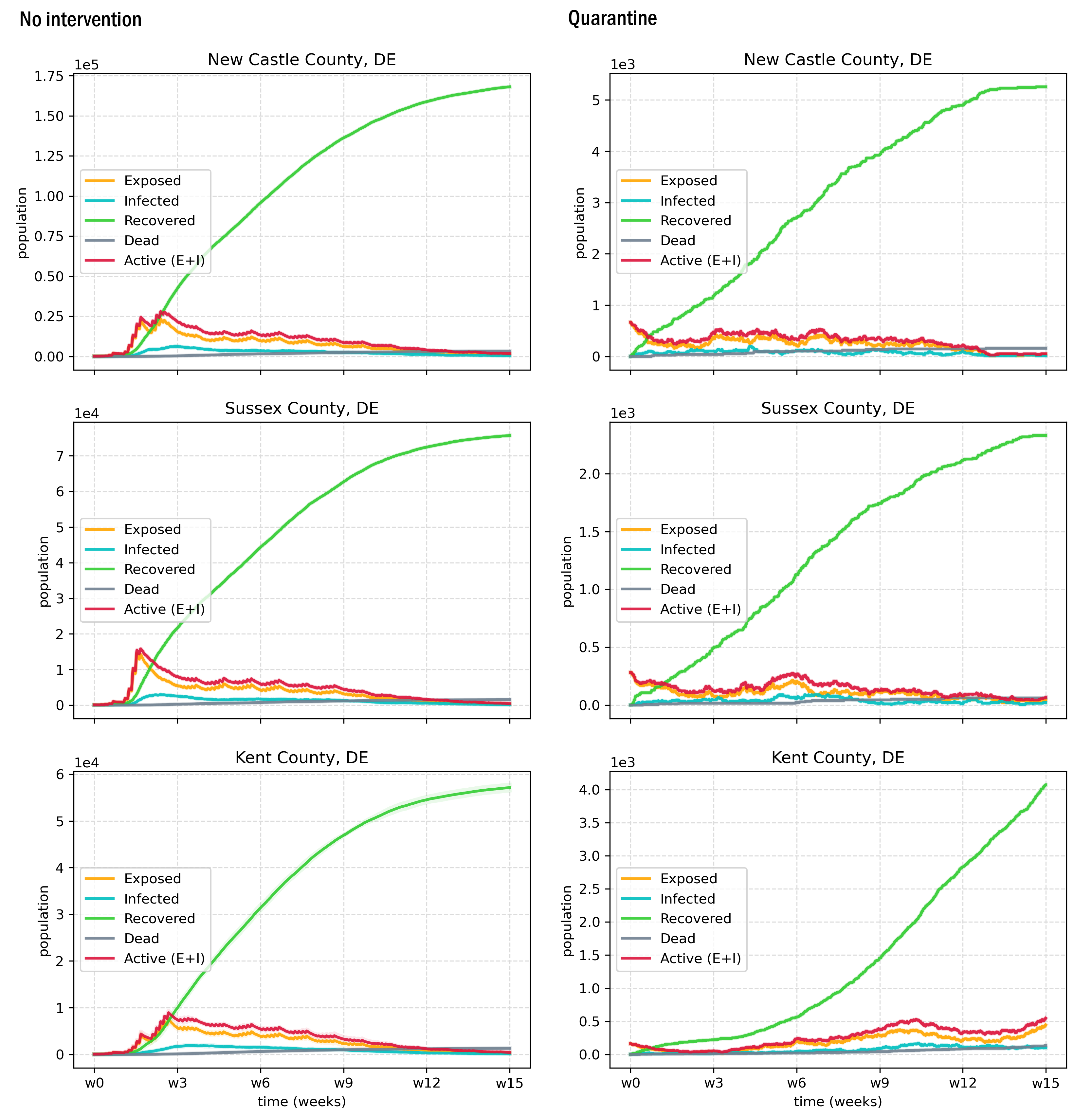}
  \caption{Occupied or needed beds in different hospitals - No intervention (left) vs. Quarantine (right)}
  \label{fig:fig14}
\end{figure}

As it is shown in Figure~\ref{fig:fig12} and Figure~\ref{fig:fig13}, the quarantine measures can drastically flatten the curves. Notice that, our quarantine scenario doesn’t impose extreme restrictions, since people can still interact with their relatives and friends and private offices, factories and crucial workplaces are still open with regular opening hours. Therefore, the results achieved emphasize that even small actions can play a prominent role.\\
Table~\ref{tab:t6} shows the number of deaths in different age groups after 15 weeks of simulation and demonstrates how vulnerable old people are and how the interventions can save their lives.\\
\begin{table}
 \caption{Average number of dead agents after 15 weeks with different measurements}
  \centering
  \begin{tabular}{lllll}
    \toprule
    \multirow{3}{*}{\textit{Age Group}} & \multicolumn{4}{c}{\textit{New Castle County}} \\
    {} & \multicolumn{2}{c}{\textit{No intervention}} & \multicolumn{2}{c}{\textit{Quarantine}} \\
    {} & \textit{Deaths} & \textit{\%of the age group} & \textit{Deaths} & \textit{\%of the age group}\\
    \midrule
    < 5 & 0.0 & 0.000\%  & 0.0 & 0.000\%\\
	5 - 17 & 20.4 & 0.022\% & 0.0 & 0.000\%\\
	18 - 44 & 273.4 & 0.132\% & 13.4 & 0.006\%\\
	45 - 64 & 911.0 & 0.645\% & 27.2 & 0.019\%\\
	65 + & 2085.8 & 2.350\% & 119.2 & 0.134\%\\
	Total & 3290.6 & 0.5889\% & 159.8 & 0.0286\%\\

	\midrule
    \multirow{3}{*}{\textit{Age Group}} & \multicolumn{4}{c}{\textit{Sussex County}} \\
    {} & \multicolumn{2}{c}{\textit{No intervention}} & \multicolumn{2}{c}{\textit{Quarantine}} \\
    {} & \textit{Deaths} & \textit{\%of the age group} & \textit{Deaths} & \textit{\%of the age group}\\
    \midrule
    < 5 & 0.0 & 0.000\%  & 0.0 & 0.000\%\\
	5 - 17 & 0.0 & 0.000\% & 0.0 & 0.000\%\\
	18 - 44 & 49.4 & 0.070\% & 0.0 & 0.000\%\\
	45 - 64 & 330.2 & 0.562\% & 2.8 & 0.005\%\\
	65 + & 1401.6 & 2.433\% & 50.6 & 0.088\%\\
	Total & 1781.2 & 0.7604\% & 53.4 & 0.0228\%\\
	
	\midrule
    \multirow{3}{*}{\textit{Age Group}} & \multicolumn{4}{c}{\textit{Kent County}} \\
    {} & \multicolumn{2}{c}{\textit{No intervention}} & \multicolumn{2}{c}{\textit{Quarantine}} \\
    {} & \textit{Deaths} & \textit{\%of the age group} & \textit{Deaths} & \textit{\%of the age group}\\
    \midrule
    < 5 & 0.0 & 0.000\%  & 0.0 & 0.000\%\\
	5 - 17 & 6.6 & 0.022\% & 0.0 & 0.000\%\\
	18 - 44 & 52.2 & 0.078\% & 4.2 & 0.006\%\\
	45 - 64 & 303.0 & 0.668\% & 3.2 & 0.007\%\\
	65 + & 916.8 & 3.130\% & 9.8 & 0.033\%\\
	Total & 1278.6 & 0.7072\% & 17.2 & 0.0096\%\\
    \bottomrule
  \end{tabular}
  \label{tab:t6}
\end{table}
The reports from Delaware, Division of Public Health \cite{DelawareDivisionofPublicHealth2020}, state that the first positive case has been identified in March 10th, 2020. As of today, after 12 weeks of spread, COVID-19 casualties raised to 149, 126 and 56 cases, for the New Castle, Sussex and Kent counties, respectively. These reports show that Sussex has a higher percentage of deaths, which matches our “no intervention” scenario. It is worth mentioning that, the infection process can be highly path-dependent when the ratio between the number of seeds and the overall population is infinitesimal. In such cases, the number of deaths can significantly vary in each iteration, and relies on the selection of the seed sets and on their actions during their infectious period. For example, if there are only few seeds with too few interactions, or they are in good health conditions to get recovered speedily, the infection won’t sprout up and won’t last for a long period.\\
Another important issue during the outbreak is the saturation of health facilities. Figure~\ref{fig:fig14} illustrates the number of beds required for the patients in each of the hospitals in the three counties. During the quarantine scenario this demand is less urgent since the speed of the virus spread slows down. The plotted lines are the cumulative of both COVID-related and non-COVID hospitalization needs.\\
\subsection{Identifying the hubs of infection}
\label{sec:hubs}
\begin{figure}
  \centering
  	\includegraphics[totalheight=9cm]{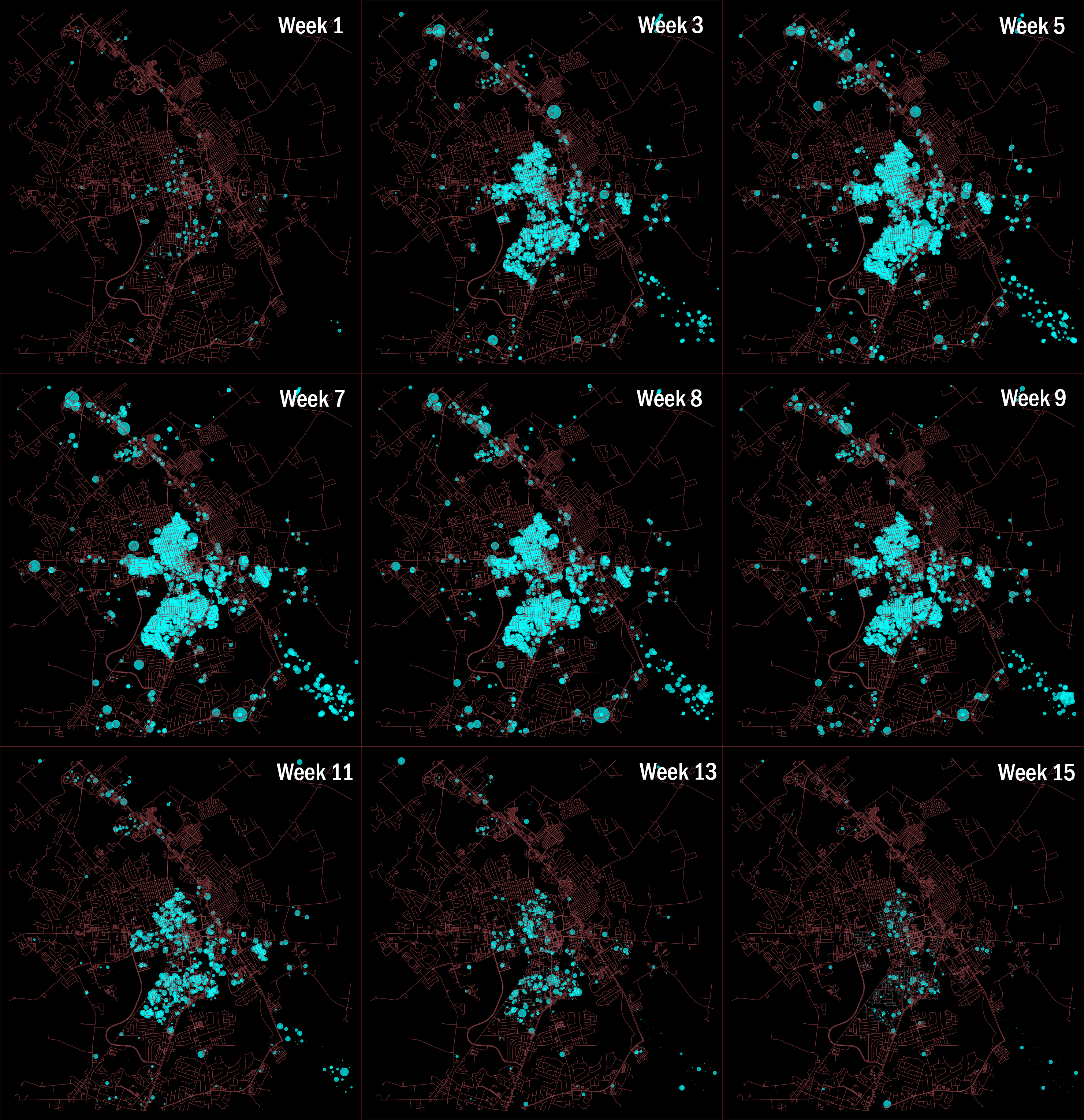}
  \caption{Simulating the spread of COVID-19 in Dover, DE with no intervention. The size of the circles represents the locations’ virus density}
  \label{fig:fig15}
\end{figure}

\begin{figure}
  \centering
  	\includegraphics[totalheight=9cm]{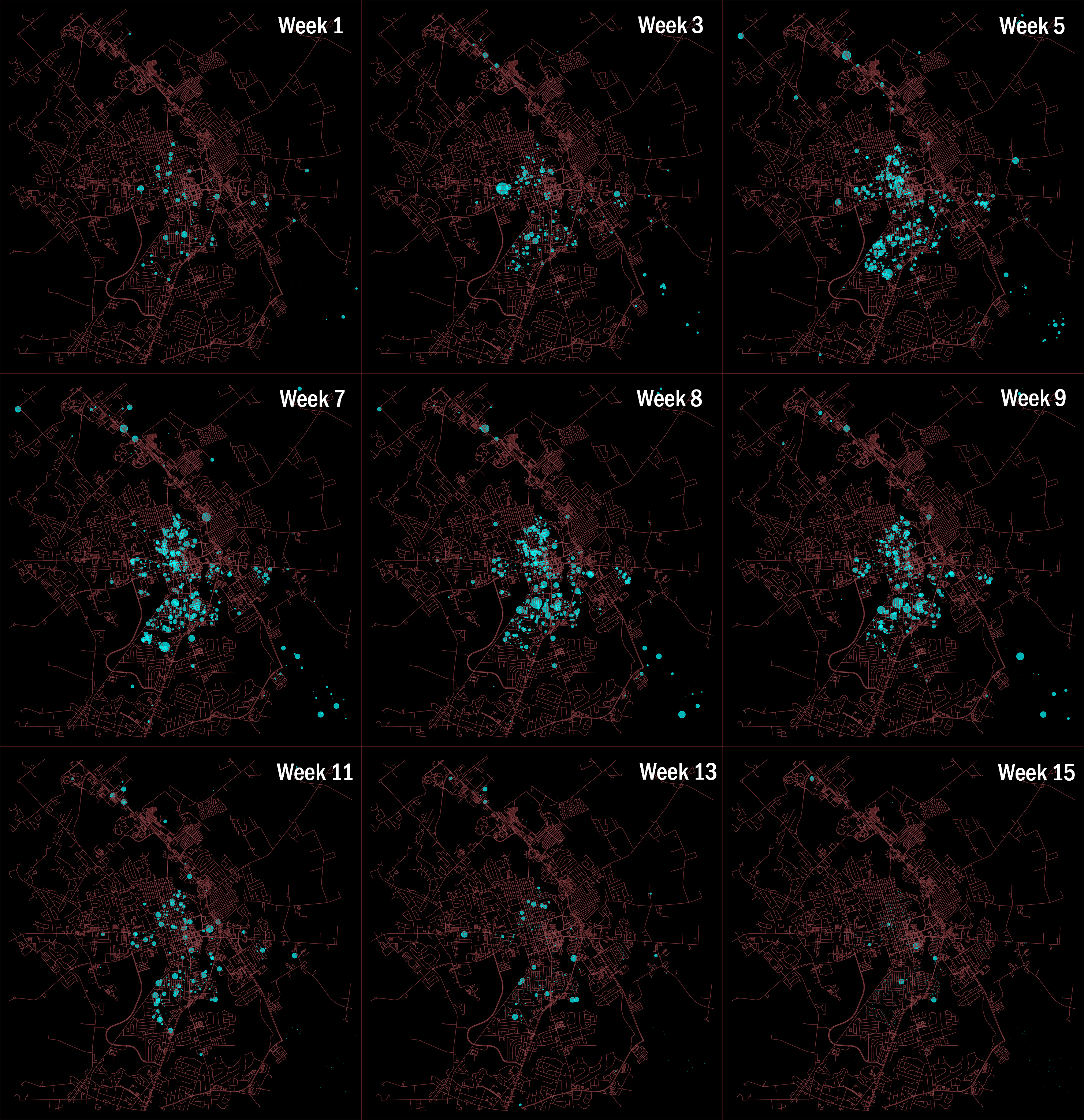}
  \caption{Simulating the spread of COVID-19 in Dover, DE in quarantine. The size of the circles represents the locations’ virus density}
  \label{fig:fig16}
\end{figure}

Our simulator has the power to monitor the spread of the virus and identify hazardous places where the probability of transmission rises. To show the phenomenon with a closer lens, we separately ran the model on the city of Dover, the capital of Delaware, with 38,000 agents. Figure~\ref{fig:fig15} and \ref{fig:fig16} depict the infected places through the time. The plots were taken at the end of every 2 weeks and illustrate the growth and decline of the outbreak in the city.\\
In order to identify the places with higher transmission potential, we extracted the areas where each “exposed” agent received the virus and became exposed. For the “no intervention” scenario, the place types inside which at least 100 incidents occurred are listed in Table~\ref{tab:t7}. Educational places collectively incubated more than half of all the virus transmissions, with approximately 7500 incidents on average. It must be pointed out that most of these cases involved young and healthy people; therefore, only a small portion of them became symptomatic infectious. On the contrary, these exposed young people conveyed the virus to their houses and made their family members exposed as well. This is one of the reasons why the number of virus transmissions inside the houses was high on average (with almost 3800 cases).\\
For the “quarantine” scenario, on the other hand, the city witnessed a dramatic decrease in the number of transmissions inside the houses due to the lock down of the educational and cultural places. Despite houses remain among the most risky locations, the average number of transmissions therein plunged to one-fifth.\\
According to Table~\ref{tab:t8}, in the quarantine scenario hospitals were the most hazardous places in terms of virus transmission. It should be emphasized that, an agent with poor health conditions or disabilities needed an eligible family member to accompany it to the hospital. On such occasion, it is very likely that the attendant gets exposed in the hospital, if he/she did not inhale the virus at home or anywhere else. Also public transportation had an important role in the viral spread, since in both scenarios it led to more than 800 incidents, on average, with a small standard deviation.\\
 \begin{table}
 \caption{The places with the most transmission incidents (No intervention)}
  \centering
  \begin{tabular}{m{4cm} m{1cm} m{2cm}}
    \toprule
    \multirow{2}{*}{\textit{Location type}} & \multicolumn{2}{c}{\textit{Number of incidents}} \\
    {} & \textit{Mean} & \textit{Standard deviation(\%)}\\
    \midrule
    Schools and Kindergartens & 5914.8 &‌ 4.2\%\\
    Houses & 3792.0 & 6.6\%\\
Universities and Colleges & 1596.0 & 3.7\%\\
Churches & 1127.2 & 4.5\%\\
Hospitals & 1012.6 & 2.5\%\\
Public transportations & 878.2 & 3.2\%\\
Gyms & 345.4 & 8.0\%\\
Café and Restaurants & 269.2 & 20.4\%\\
Museums & 201.0 & 20.3\%\\
Private Offices & 154.8 & 1.0\%\\
Industrial plants & 150.4 & 23.4\%\\
    \bottomrule
  \end{tabular}
  \label{tab:t7}
\end{table}
\begin{table}
 \caption{The places with the most transmission incidents (Quarantine)}
  \centering
  \begin{tabular}{m{4cm} m{1cm} m{2cm}}
    \toprule
    \multirow{2}{*}{\textit{Location type}} & \multicolumn{2}{c}{\textit{Number of incidents}} \\
    {} & \textit{Mean} & \textit{Standard deviation(\%)}\\
    \midrule
Hospitals & 934.0 & 0.8\%\\
Public transportations & 802.2 & 8.3\%\\
Houses & 687.0 & 3.2\%\\
Groceries & 37.2 & 24.0\%\\
Private Offices & 29.2 & 9.1\%\\
Industrial plants & 28.0 & 3.6\%\\
Shops & 13.6 & 23.5\%\\
Banks & 13.4 & 22.9\%\\
Gas stations & 11.0 & 39.6\%\\

    \bottomrule
  \end{tabular}
  \label{tab:t8}
\end{table}

The comparison between Figure~\ref{fig:fig15} and \ref{fig:fig16} graphically demonstrates the effectiveness of the quarantine measures in terms of the outbreak mitigation. During the quarantine, indeed, the prevalence and density of the virus inside the city buildings and the outdoor areas are considerably lower with respect to the other scenario.\\
Moreover, both figures show that the buildings in the city downtown (the center of the map) can reach the highest virus densities. In particular, Figure~\ref{fig:fig15} illustrates that the areas with a higher number of churches, colleges and schools (e.g. the downtown), the south-eastern district and the northern district, where Delaware State University and Wilmington University are located, have higher likelihoods to become virus hubs.
\section{Discussion}
\label{sec:dis}
ABM is one of the computational approaches which can simulate individuals and their decisions based on the rules and the set of actions that are programmed in it. These rules are configured so that the simulator can better mimic the behaviour of its real-world counterpart. Such simulators can be of great importance when the interactions among the citizens and their decisions profoundly shape the dynamics of the system. Thus, since the intensity of the outbreaks is highly reliant upon the behaviour of the individuals, this paper proposes an ABM framework, named CoV-ABM, considering such actions and decisions in a higher detail compared to other existing frameworks. In addition, due to the unique epidemiological and clinical features of the novel coronavirus SARS-CoV2 \cite{Wang2020}, and its remarkable durability and persistence in the environment \cite{Kampf2020}, the development of a simulator being able to consider these characteristics was needed.\\
During the past few months, the COVID-19 pandemic heavily affected the lives of billions of people around the world, as well as the global economy. Different countries and international organizations are taking unprecedented measures to combat its spread \cite{InternationalLabourOrganization2020}. In this scenario, CoV-ABM can be used as a decision-support tool for policymakers and researchers to analyse the consequences of different decisions and regulations devoted to halt the outbreak and prevent future potential occurrences.\\
This paper proposes a novel extension of SEIR, called SEIHRD, which formulates the hospitalized cases and the saturation of the facilities and considers both symptomatic and asymptomatic cases. According to a study from China \cite{Bai2020}, approximately 80\% of SARS-CoV2 cases had no clinical symptoms, even though they could release large amounts of viruses at the early phases of the infection. CoV-ABM was developed to being able of tracking the active viruses existing at every indoor or outdoor areas, and computing the transmission probability based on the virus density at each location. In the proposed model, indeed, a place visited by infectious people remains infectious until all its viruses get inactivated.\\
This novel assumption, unlike that of studies whose diffusion models rely on social networks or agent-agent interactions, is totally compatible with the peculiarity of SARS-CoV2. Former studies assumed that an agent directly transmits the virus to another agent when both are close enough for a certain period or time, or when a graph-edge gets activated. In our model, instead, an agent can infect any other individual indirectly  through the environment. As an example, agent $a^{v_{i},u_{j}}$ releases an amount of virus at location y, where it passes by $y$ at time $t$. Any susceptible agent $a^{v_{m},u_{n}}$ who visits the same location at time t might get exposed regardless of having a social link with agent $a^{v_{i},u_{j}}$, until the viruses in that area are not disabled (i.e. $y^{\Delta^{t}}>0$). In graph-based models this can be possible only if the graph is complete.\\
Another novelty of this research is how the agent communities and their interactions were modelled, specifically the interactions among family members, relatives and friends. In our framework, each agent is a member of a family, has a group of relatives and a set of friendship groups. When a group of friends or relatives decide to visit each other, they choose a destination among the possible open locations where they can go together. Even if no open public places exist, they still have the chance to set the destination to one of the available members’ houses; this possibility enables the escalation of the virus spread. Hence, an effective mitigating scenario should consider not only limitations of the opening hours of public places but also measures devoted to lowering as possible the probability of interactions among individuals.\\
By defining two scenarios, “no intervention” and “quarantine”, this study aims at demonstrating how lock down measures can halt the outbreak and “flatten the curve”. In particular, the simulation results for the State of Delaware showed that the restrictions can hugely reduce the number of infected people and, consequently, the number of victims and the hospital bed demand. From this it can be inferred that in the “no intervention” scenario, Delaware hits the peak of outbreak between the 2nd and the 3rd week, and the number of active persons remains relatively high for about 9 more weeks. In the “quarantine” scenario, instead, the outbreak peak happens somewhere between the 6th and the 12th weeks while the number of active agents is relatively low and lasts without any significant change by the end of our simulation period. The results also illustrated that the spread dynamics slightly differ in different counties, maybe due to their different health structures and urban arrangements. This opens up an opportunity for future studies to investigate the impact of urban design and street graphs on the spread of COVID-19.\\
By tracking the locations of all transmission incidents and calculating the virus densities for different areas, CoV-ABM highlighted the high-risk places inside the geographical environment. For the city of Dover, the simulation results showed how educational and cultural places can become hubs of infection that accelerate the outbreak. In the “no intervention” scenario, educational buildings caused roughly 48\% of all the transmissions. The same percentage for cultural places, including churches and museums, was just over 8\% on average. On the other hand, our results draw attention on the protection policies inside hospitals and public transportation, since the simulations showed that these places, taken together, can cause more than 70\% of all the infection incidents during the quarantine. Households achieved a share of 25\% and 23\% of all the incidents for the “quarantine” and the “no intervention” scenarios, respectively.\\
Notice that, this paper tried to gather and use all the available real-world data to generate a simulation environment as accurate as possible. However, there was still a substantial portion of urban and suburban buildings for which complete information was missing, such as their usage, their indoor volume size and their opening hours. In such cases, CoV-ABM assigned random values to make the environment complete. In order to enhance the precision of the simulator, more comprehensive census and geographical information databases are required.
\section{Conclusion and future opportunities}
\label{sec:conclusion}
In this paper we proposed a novel spatiotemporal simulation framework, called CoV-ABM, to mimic the spread of SARS-CoV2 virus in different geographical domains. This framework uses census data and GIS information to generate a realistic geographical environment where people live, work, and interact with each other. CoV-ABM employs a new extension of SEIHRD diffusion model which considers the virus characteristics and its ability to persist in the air and/or on surfaces for long periods of time. Our diffusion model works based on the density of active SARS-CoV2 viruses at each place’s volume. Therefore, unlike other simulators proposed in the literature, agents in our framework can inhale the virus in an infectious place, even if there is no other agent around. Moreover, the diffusion model considers also the actual number of infectious people who can have asymptomatic periods of illness, and not only those who have been tested, and some risk factors, such as pregnancy, smoking record, heart diseases and kidney misfunctioning, to generate the health profiles for each population based on the local health information.\\
CoV-ABM simulates the agents’ daily needs and their social interactions with a high resolution, which enabled us to study the consequences of different mitigation policies. Therefore, in this study we examined the impact of two different scenarios on the dynamics of the outbreak in the State of Delaware, with a population of one million. The simulation results illustrated that the number of deaths drastically grew in different age-groups in case of no interventions compared to the quarantine scenario. Our framework assumed that the plethora of agents inside hospitals and the higher number of hospitalizations reduced the hospital efficiency, and compared the difference between the hospitals’ bed demands in the two scenarios.\\
Finally, CoV-ABM identified the high risky locations and buildings, in terms of the virus transmission likelihood, in the geographical areas under investigation. According to the simulation results, the top three virus hubs were schools, houses and universities when no restrictions were imposed, while the top-ranked places were hospitals, public transportations and houses in the quarantine scenario. These outcomes can be of great importance for policymakers to understand the consequences of their decisions, in particular of laws and protective regulations focused on limiting the attendance of such locations.\\
The current implementation of CoV-ABM was run by simulating millions of agents  at a country-level geography, to control the computational effort and enable its usage on a personal computer. The proposed simulator, however, could be applied on continental domains and/or on a world-wide scale by adding or removing few assumptions.
Several research opportunities and working directions are identified and recommended as follows: (1) developing a variety of different scenarios based on the current or future policy responses to COVID-19; (2) extending the SEIHRD model to separate the ICU beds and regular beds for patients with different morbidity conditions at hospitals and considering the effect of self-quarantine; (3) adding airports and harbours to the model and investigating the impact of travellers and tourists on the spread dynamics;(4) investigating the influence of different urban designs on the severity of outbreak.
\section*{Acknowledgement}
This research did not receive any specific grant from funding agencies in the public, commercial, or not-for-profit sectors.

\section*{CRediT author statement}
\label{sec:credit}
\textbf{Masoud Jalayer}: Methodology, Conceptualization, Software, Data Curation, Investigation, Visualization, Writing- Original draft preparation. \textbf{Carlotta Orsenigo}: Validation, Writing- Reviewing and Editing, Supervision. \textbf{Carlo Vercellis}: Supervision\\
\newpage
\appendix
\section{Algorithms}
\label{app:alg}

\begin{algorithm}[H]
 \caption{Pseudo-code of the SEIHRD model}
 \label{alg:seir}
 \KwData{$t, a^{v,u} , locations, parameters$}
 \KwResult{$a^{v,u} , locations$}
 \uIf{$a^{v,u}(s^{t})=infectious$}{
  infect $a^{v,u}(\varrho^{t})$ ; calculate $a^{v,u}(\phi^{t+dt})$ \;
  \If{$a^{v,u}(\varrho^{t})\neq a^{v,u}(\varrho^{t+dt})$}{
   infect $a^{v,u}(\varrho_{out}^{t})$ and $a^{v,u}(\varrho_{out}^{t+dt})$\;   }
   \eIf{$a^{v,u}(\varrho^{t}) \in hospitals$}{
    calculate $H^t_{q}$\;
	    \uIf{random<$p_{h\rightarrow d}^{t}$}{
    set $a^{v,u}(s^{t}) \leftarrow dead$; set $a^{v,u}(\varrho^{t+dt}) \leftarrow$ a cemetery nearby}
        \ElseIf{random<$p_{h\rightarrow r}^{t}$}{
    set $a^{v,u}(s^{t}) \leftarrow recovered$;}
   }
   {\uIf{random<$p_{i\rightarrow d}^{t}$}
   {set $a^{v,u}(s^{t}) \leftarrow dead$; set $a^{v,u}(\varrho^{t+dt}) \leftarrow$ a cemetery nearby}
   {\ElseIf{random<$p_{i\rightarrow r}^{t}$}
    {set $a^{v,u}(s^{t}) \leftarrow recovered$;}
   }
  }
  }
  {\uElseIf{$a^{v,u}(s^{t})=exposed$}
  {infect $a^{v,u}(\varrho^{t})$\;
  \If{$a^{v,u}(\varrho^{t})\neq a^{v,u}(\varrho^{t+dt})$}{
   infect $a^{v,u}(\varrho_{out}^{t})$ and $a^{v,u}(\varrho_{out}^{t+dt})$\;
   \uIf{random<$p_{e\rightarrow r}^{t}$}{
    set $a^{v,u}(s^{t}) \leftarrow recovered$;}
    {\ElseIf{random<$p_{e\rightarrow i}^{t}$}
    {set $a^{v,u}(s^{t}) \leftarrow infectious$;}}
   }}
  }
  {\ElseIf{$a^{v,u}(s^{t})=susceptible$}
  {\If{random<$p_{s\rightarrow e}^{t}$}
  {set $a^{v,u}(s^{t}) \leftarrow exposed$;}
  }}
\end{algorithm}
\newpage
\begin{algorithm}[H]
 \caption{Pseudo-code of the CoV-ABM discrete-event model}
 \label{alg:covabm}
 \textbf{Initialization:} set objects to default values; \textbf{for} $(v,u) \in seeds$ \textbf{then} set $a^{v,u}(s^{0})\leftarrow exposed$\;
 \textbf{Main loop:}\\
 \While{$t<t^{max}$}{
	\ForEach{$\varrho \in$ locations}{update $\varrho(\Delta^{t})$}
	\ForEach{$v \in$ families}{
		\ForEach{$n \in$ daily needs}{update $v_{n}^{t}$}\
		\ForEach{$u \in v(members)$}{update $v_{emergency}^{regular}(u)^{t}$\;
		\If{$v_{emergency}^{regular}(u)^{t}=1$}{set $a^{v,u}(\varrho^{t+dt}) \leftarrow$ a related place nearby; set $v_{emergency}^{regular}(u)(t_{0})\leftarrow t+dt$}
		update $a^{v,u}(\phi^{t+dt})$;
		update $a^{v,u},locations \leftarrow SEIHRD(t, a^{v,u} , locations, parameters)$

	\If{$a^{v,u}(s^{t})=infectious$ \& $a^{v,u}(\varrho^{t})\notin hospitals$}{update $p_{i\rightarrow h}^{t}$}
	\If{$random<p_{i\rightarrow h}^{t}$}{update $v_{emergency}^{COVID}(u)^{t}$; set $a^{v,u}(\varrho^{t+dt})\leftarrow$ a related emergency place nearby}
	\ForEach{$n \in needs=\lbrace n\vert v_{n}^{t}\neq \varnothing \rbrace$}{\textbf{if} $a^{v,u}$ is eligible \textbf{then} set $O_{n}^{t+dt} \leftarrow \lbrace o_{i}\vert \Vert shortest path(o_{i},a^{v,u}(\varrho^{t})\Vert <2km\rbrace$\;
	\If{$O_{n}^{t+dt} \neq \varnothing$}{set $a^{v,u}(\varrho^{t+dt})\leftarrow$ random choice from $O_{n}^{t+dt}$; set $v_{n}^{t}\leftarrow \varnothing$; set $v_{n}^{t+dt}\leftarrow \varnothing$;}
		}
	\If{$\lbrace v_{emergency}^{COVID}(u)^{t}\bigcup v_{emergency}^{regular}(u)^{t}\rbrace \neq \varnothing$}{\If{$a^{v,u}(\zeta)<18$ or $a^{v,u}(h)<0.3$ or $a^{v,u}(d)=1$}{$a^{v,q} \leftarrow$ random choice from eligible family members (if any) for help\;
	set $a^{v,q}(\varrho^{t+dt})\leftarrow a^{v,u}(\varrho^{t+dt})$}}
	\uIf{$v_{emergency}^{COVID}(u)^{t} \neq \varnothing$}{\textbf{if} $a^{v,u}(s^{t})\neq infectious$ \textbf{then} set $v_{emergency}^{COVID}(u)^{t} \leftarrow \varnothing$}
	\ElseIf{$t-v_{emergency}^{regular}(u)(t_{0})\geq L$}{set $v_{emergency}^{regular}(u)^{t} \leftarrow \varnothing$}
	}}
	\ForEach{$g \in friend groups$}{set $F\leftarrow \lbrace (v,u)\vert a^{v,u}(\phi^{t+dt})=1\rbrace$\;
	\If{$random<p_{friends}\times time factor(t+dt)$ and $\Vert F\Vert>1$}{set $l\leftarrow$ a random location where they can get together\;
	\ForEach{$(v,u)\in g$}{set $a^{v,u}(\varrho^{t+dt})\leftarrow l$; set $a^{v,u}(\phi^{t+dt})\leftarrow 0$}
	}
	}
	\ForEach{$g \in relatives groups$}{\ForEach {$v \in g$}{set $\Phi^{v}\leftarrow \lbrace (v,u)\vert a^{v,u}(\phi^{t+dt})=1\rbrace$}
	set $F_{g}\leftarrow \lbrace v_{i} \vert \Vert \Phi^{v_{i}}\Vert>1\rbrace$\;
	\If{$random<p_{relatives} \times time factor(t+dt)$ and $\Vert F_{g}\Vert>1$}{set $l\leftarrow$ a random location where they can get together\;
	\ForEach{$v\in g$}{\ForEach{$u \in \Phi^v$}{set $a^{v,u}(\varrho^{t+dt})\leftarrow l$; set $a^{v,u}(\phi^{t+dt})\leftarrow 0$}}}
	}
	\ForEach {$h \in hospitals$}{update $H_h^t$}
	set $t \leftarrow t+dt$
}

\end{algorithm}

\bibliographystyle{abbrvnat}
\bibliography{Covid}

\end{document}